\documentclass[12pt]{article}
\usepackage{amsmath,amssymb,exscale,cancel}
\usepackage{graphicx}
\usepackage{epsfig}
\usepackage{multicol}
\usepackage{color}
\usepackage{mathrsfs}
\usepackage{blindtext}
 \usepackage{fancyhdr}
\usepackage{hyperref}
\usepackage{cite}
\usepackage{mathtools}

\usepackage{floatrow}

\usepackage{graphicx}
\usepackage{sidecap}

\usepackage[latin1]{inputenc} 
\usepackage{multicol}  
 
 \usepackage{titlesec}
 
 \usepackage{rotating,slashed,xcolor,amsfonts,expdlist,charter}

\numberwithin{equation}{section}

\usepackage{xcolor}
\usepackage{sectsty}

\usepackage{mdframed}
\usepackage{titletoc}

\definecolor{secnum}{RGB}{13,151,225}
\definecolor{ptcbackground}{RGB}{212,237,252}
\definecolor{ptctitle}{RGB}{0,177,235}

\titlecontents{lsection}
  [5.8em]{\sffamily}
  {\color{secnum}\contentslabel{2.3em}\normalcolor}{}
  {\titlerule*[1000pc]{.}\contentspage\\\hspace*{-5.8em}\vspace*{5pt}%
    \color{white}\rule{\dimexpr\textwidth-15.5pt\relax}{1pt}}

\usepackage{hyperref}
\hypersetup{colorlinks,bookmarksopen,bookmarksnumbered,citecolor=rossos,
linkcolor=redy,pdfstartview=FitH,urlcolor=blus}
\usepackage{slashed}

\definecolor{blus}{cmyk}{1,1,0,0.1}
\definecolor{verdes}{cmyk}{0.99,0,0.59,0.65}
\definecolor{rossos}{cmyk}{0,1,1,0.55}
\definecolor{redy}{cmyk}{0,1,1,0.7}
\definecolor{greeny}{cmyk}{0.99,0,0.59,0.98}
\definecolor{green-go}{cmyk}{0.79,0,0.59,0.5}

\usepackage{titlesec}

\def\Lag{\mathscr{L}}

\newcommand{\beq}{\begin{equation}}
\newcommand{\eeq}{\end{equation}}

\def\hhref#1{\href{http://arxiv.org/abs/#1}{arXiv:#1}} 

 \def\Lag{\mathscr{L}}
 
\newcommand{\tmtextbf}[1]{{\bfseries{#1}}}
\newcommand{\tmtextrm}[1]{{\rmfamily{#1}}}
\newcommand{\bp}{\bar M_P}

\def\be{\begin{equation}}
\def\ee{\end{equation}}
\def\ba{\begin{array} }
\def\bac{\begin{array} {c}}
\def\bacc{\begin{array} {cc}}
\def\baccc{\begin{array} {ccc}}
\def\bacccc{\begin{array} {cccc}}
\def\ea{\end{array}}
\def\bea{\begin{eqnarray}}
\def\eea{\end{eqnarray}}

\definecolor{red}{rgb}{1,0,0}

\def\psl{\hbox{\hbox{${p}$}}\kern-1.9mm{\hbox{${/}$}}}
\def\dsl{\hbox{\hbox{${\partial}$}}\kern-2.2mm{\hbox{${/}$}}}
\def\Dsl{\hbox{\hbox{${D}$}}\kern-2.6mm{\hbox{${/}$}}}

\def\Lag{\mathscr{L}}

\newcommand{\gappeq}{{\rlap{{\raise}.5ex\text{\ensuremath{>}}}{{\lower}.5ex\text{\ensuremath{\sim}}}}}
\newcommand{\lappeq}{{\rlap{{\raise}.5ex\text{\ensuremath{<}}}{{\lower}.5ex\text{\ensuremath{\sim}}}}}

\newcommand{\I}{\tmtextrm{1{\kern}-.24em l}}

\newcommand{\gt}{f_2}
\newcommand{\dxi}{\zeta}

\begin{document}
\topmargin -1.0cm
\oddsidemargin 0.9cm
\evensidemargin -0.5cm

{\vspace{-1cm}}
\begin{center}

\vspace{-1cm}

 {\Huge \tmtextbf{ \color{blus} 
 Quasi-Conformal Models \\ and the Early Universe}} {\vspace{.5cm}}\\
 
\vspace{0.9cm}

{\large  {\bf Alberto Salvio }
\vspace{.3cm}

{\em  

\vspace{.4cm}
Physics Department, University of Rome and INFN Tor Vergata, Italy\\ 

\vspace{0.4cm}

\vspace{0.2cm}

 \vspace{0.5cm}
}}

\noindent ------------------------------------------------------------------------------------------------------------------------------

\end{center}

\begin{center}
{\bf \large Abstract}
\end{center}

\noindent Extensions of the Standard Model and general relativity featuring a UV fixed point can leave observable implications at accessible energies. Although mass parameters such as the Planck scale can appear through dimensional transmutation, all fundamental dimension-4 operators can (at least approximately) respect Weyl invariance at finite energy. An example is the Weyl-squared term, whose consistency and observational consequences are studied. This quasi-conformal scenario emerges from the UV complete quadratic gravity and is a possible framework for  inflation.   We find two  realizations. In the first one the inflaton is a fundamental scalar with a quasi-conformal non-minimal coupling to the Ricci scalar.  In this case the field excursion must not exceed the Planck mass by far. An example discussed in detail is hilltop inflation. In the second realization the inflaton is  a pseudo-Goldstone boson (natural inflation). In this case we show how to obtain an elegant UV completion within an asymptotically free QCD-like theory, in which the inflaton is a composite scalar due to new strong dynamics. We also show how efficient reheating can occur.  Unlike the natural inflation based on Einstein gravity, the tensor-to-scalar ratio is well below the current bound set by Planck. In both realizations mentioned above, the basic inflationary formul\ae\,\,are computed analytically  and, therefore, these possibilities  can be used as simple benchmark models.

  \vspace{0.9cm}

\noindent -------------------------------------------------------------------------------------------------------------------------------

\vspace{0.5cm}

\vspace{2cm}

\noindent Email: alberto.salvio@roma2.infn.it 

\vspace{0.cm}
%
\noindent 


\vspace{-.5cm}

\noindent 
 \newpage

\tableofcontents

\vspace{0.2cm} 



\section{Introduction}

 In 2015 (and with a recent update in 2018) the Planck collaboration~\cite{Ade:2015lrj,Akrami:2018odb} was able to exclude (or set stringent bounds on) several inflationary models. Indeed, better determinations of many observables related to the cosmic microwave background (CMB) (such as the tensor-to-scalar ratio $r$, the scalar spectral index $n_s$ and the curvature power spectrum $P_R$) were provided. Further improvements are expected in the next future: CMB Stage 4 (S4) will be active soon\footnote{See, for example, the   webpage \href{https://cmb-s4.org/overview.php}{https://cmb-s4.org/overview.php}.}. Therefore, early universe cosmology continues to be an exciting research area.
 
 Despite this  big progress, several models of the early universe are still allowed and we have only a partial understanding of how the universe initially expanded and evolved. Such large ambiguity can be reduced by looking for theoretical reasons to discriminate among the various possibilities. One way to do so is to focus on UV-complete theories that are applicable to the early universe.
 
 In Ref.~\cite{Salvio:2017qkx} it was shown how a relativistic field theory of all interactions (gravity included) can reach infinite energy. In order to achieve this goal, terms quadratic in the curvature are added to the Einstein-Hilbert action, so this theory is also known as quadratic gravity (QG) (see~\cite{Salvio:2018crh} for a review). The UV completion is obtained by demanding that the Weyl symmetry breaking terms vanish in the infinite energy limit and all couplings enjoy a UV fixed point. Therefore, this is a way of implementing asymptotic safety.
An important difference between this possibility and the original proposal made by Weinberg in~\cite{WeinbergAS}, though, is the presence of only a finite number of terms in the fundamental action, which guarantees the theory to be  predictive and calculable. Another interesting property of QG is the possibility to solve the hierarchy problem~\cite{Salvio:2014soa,Salvio:2017qkx}, in the sense that the Higgs mass can be made technically natural~\cite{tHooft:1979rat}. When one lowers the energy from the UV-conformal fixed point mentioned above,  the dimensionless Weyl-symmetry breaking operators are generated, but since they are sourced only by multiloop diagrams~\cite{Hathrell:1981zb,Hathrell:1981gz,Jack:1990eb}, they can remain small down to the inflationary energies. This corresponds to a scenario for the early universe, which we call ``quasi-conformal"\footnote{Another independent motivation to study this scenario comes from string theory, where many  scalars are (quasi-)conformally coupled to gravity~\cite{Kofman:2007tr}.}. 

One of the quadratic terms in the action is the squared of the Weyl tensor, which is necessary to keep the theory UV complete.  Featuring higher time derivatives, this  Weyl-squared term brought a number of issues, which, however, have been addressed in the recent years. One of the purposes of this paper is to review and extend these arguments in favour of the viability of such term.

Another aspect that has not been investigated  so far is the observational  predictions of the quasi-conformal scenario. The natural arena to study these effects is the early universe, which typically involves extremely high energies and field scales, sometimes reaching the Planck mass $\bp$. This is the main topic of the present paper\footnote{See Refs.~\cite{Kobayashi:2008rx,Buck:2010sv,Boran:2017cfj} for other  related studies.}.

The ``quasi" of ``quasi-conformal" is important. It reminds us that the Einstein-Hilbert term and other Weyl-breaking parameters can be present in the Lagrangian, generated, for example, through dimensional transmutation and various quantum effects~\cite{Salvio:2014soa,Zee:1980sj,Adler:1982ri,Kannike:2015apa,Salvio:2017qkx,Holdom:2015kbf,Donoghue:2016xnh,Donoghue:2017vvl,Donoghue:2018izj,Karam:2018mft,Kubo:2018kho}. Note that an exactly Weyl invariant model would not be physically different from a Weyl-breaking one; indeed, any exactly conformal model can be written as
a non-conformal one by fixing a gauge for Weyl symmetry and, conversely, any model of gravity coupled to an arbitrary matter sector can always be made exactly Weyl invariant\footnote{This can be achieved  by simply writing the spacetime metric $g_{\mu\nu}(x)$ 
as $S(x) \bar g_{\mu\nu}(x)$, where $S$ is a scalar quantity that depends on the spacetime point $x$ and $\bar g_{\mu\nu}$ is a redefined metric. Indeed, this introduces the gauge redundancy $S(x)\to \lambda (x) S(x)$, 
$\bar g_{\mu\nu}(x)\to \bar g_{\mu\nu}(x)/\lambda(x)$, where $\lambda(x)$ is a generic function of $x$ (in other words, it makes the model exactly Weyl invariant).}. What is proposed here is quasi-conformality, in which Weyl symmetry is approximate at high energy, but not exact apart from the strict UV limit. This is also consistent with the presence of the Weyl anomaly, which would anyhow render exact conformality impossible (when the theory is not at the fixed point, for instance at accessible energies).

A first question one can ask is whether the Standard Model (SM) Higgs can play the role of the inflaton in this scenario. The answer is  negative because the requirement of a quasi-conformal non-minimal coupling (to the Ricci scalar) leads essentially to the same problem one has for a vanishing non-minimal coupling~\cite{Isidori:2007vm}: the SM Higgs cannot be responsible for inflation and the generation of the observed perturbations at the same time. We have verified the validity of this statement for a quasi-conformal non-minimal coupling by using the precise 2-loop effective potential as in~\cite{Bezrukov:2009db,Degrassi:2012ry,Buttazzo:2013uya,Salvio:2013rja,Allison:2013uaa}. A way to circumvent this problem is to demand a large~\cite{Bezrukov:2007ep,Barvinsky:2008ia} (or order 10~\cite{Hamada:2014iga,Bezrukov:2014bra,Hamada:2014wna,Salvio:2017oyf}) non-minimal coupling, which, however, would completely  break  Weyl symmetry in the UV. 

Another apparently natural candidate for the inflaton is the effective spinless field corresponding to the square of the Ricci scalar, as proposed by Starobinsky~\cite{Starobinsky:1980te}. However, in the quasi-conformal scenario the coefficient of this term has to be very small because is not Weyl-invariant, while the observed value of the curvature power spectrum requires a very large coefficient.

In this paper we discuss two implementations of the quasi-conformal scenario. The first one is the class of models where the inflaton features a moderate field excursion: with ``moderate" we mean that it  does not exceed the Planck scale by far. This is because a quasi-conformal non-minimal coupling prohibits, as we will discuss in Sec.~\ref{model}, field excursions bigger than $\sqrt{6}\bp$. A specific realization that we study in detail here is hilltop inflation~\cite{Boubekeur:2005zm}.

Another interesting implementation that we investigate in this paper is inflation triggered by a pseudo-Goldstone boson, what is known as ``natural inflation" in the context of the non-conformal Einstein gravity~\cite{Freese:1990rb}. Indeed, in this case the inflaton does not need to a have a (quasi-)Weyl invariant non-minimal coupling being morally the {\it phase} of some field representation of a spontaneously broken global group. An attractive feature of natural inflation is the natural flatness of the  potential, which is protected by Goldstone theorem from large quantum effects. This is preserved in (quasi-)conformal realizations and we will study possible ways of distinguishing this setup from the one based on Einstein gravity.

The paper is organized as follows. In the next section we describe in full generality the quasi-conformal scenario. As anticipated before, the conformal Weyl-squared term is introduced and we review and extend arguments which indicate the viability of this term. In Sec.~\ref{model} we study the hilltop quasi-conformal model, while in Sec.~\ref{Natural inflation} we focus on the natural-inflation option, investigating how the universe inflated and subsequently evolved in this case. Finally, Sec.~\ref{conclusions} offers our conclusions and some outlook of this work.

\section{The quasi-conformal scenario and the Weyl-squared term}\label{The quasi-conformal scenario}

The quasi-conformal scenario studied in this work is a subset of the theories 
with scale invariance in the UV. Therefore, we start here by reviewing this class of theories. In the most general case, the field content features gauge fields with field strength $F_{\mu\nu}^A$, Weyl fermions $\psi_j$ and real scalars $\phi_a$. These fields should be considered as fundamental. Other composite states can appear below some confinement scales as usual.

The full action in the so-called Jordan frame is 
\be S=\int d^4x  \sqrt{-g} \, \Lag, \qquad \Lag=  \mathscr{L}_{\rm SI} + \Lag_{\cancel{\rm SI}} \label{TotAction}. \ee
where $g$ is the determinant of the metric $g_{\mu\nu}$. We describe in turn the two pieces $\mathscr{L}_{\rm SI}$ and $\Lag_{\cancel{\rm SI}}$. 

$\mathscr{L}_{\rm SI}$ is the scale invariant part.
Since we are interested in theories with an approximate Weyl symmetry in the UV we split the first term as $\mathscr{L}_{\rm SI}= \mathscr{L}_{\rm Weyl}+\Lag_{\cancel{\rm Weyl}}$, where the first part is invariant under Weyl transformations
\bea \label{eq:Lconf}
\Lag_{\rm Weyl} &=& -\frac{W^2}{2\gt^2} 
- \frac14 (F_{\mu\nu}^A)^2 + \frac{(D_\mu \phi_a)^2}{2}  + \bar\psi_j i\slashed{D} \psi_j  +\nonumber\\
&& + \frac{1}{12} \phi_a^2 R - \frac12 (Y^a_{ij} \psi_i\psi_j \phi_a + \hbox{h.c.}) - \frac{\lambda_{abcd}}{4!} \phi_a\phi_b\phi_c\phi_d
\label{eq:Lgen}\eea
and the second part
\beq  \label{eq:Lnotconf}
\Lag_{\cancel{\rm Weyl}} =  \frac{R^2}{6f_0^2}  -  \frac12 \dxi_{ab} \phi_a \phi_b R
\eeq
is not invariant. If one rewrites the coefficients $\zeta_{ab}$ as $\zeta_{ab}=\xi_{ab} +\delta_{ab}/6$ one recovers the usual notation for the non-minimal couplings $\xi_{ab}$.  Here $W^2\equiv W_{\mu\nu\rho\sigma}W^{\mu\nu\rho\sigma}$, where $W_{\mu\nu\rho\sigma}$ is the Weyl tensor. $\mathscr{L}_{\rm SI}$ contains dimensionless parameters only: the gravitational ones $f_0$ and $f_2$, the non-minimal couplings $\xi_{ab}$ and the Yukawa $Y^a_{ij}$, quartic  couplings $\lambda_{abcd}$ and the gauge couplings in the covariant derivatives $D_\mu$.

$\Lag_{\cancel{\rm SI}}$ contains instead the massive parameters and, although irrelevant in the UV, they play an important role in the IR:
\beq \label{massiveL}\Lag_{\cancel{\rm SI}} =-   \frac12 \bp^2 R - \Lambda-  \frac12 m^2_{ab} \phi_a \phi_b - \frac16 A_{abc}\phi_a\phi_b\phi_c
-\frac12 (M_{ij}\psi_i\psi_j+\hbox{h.c.}), 
\eeq
where $\bp$ is the reduced Planck scale, $\Lambda$ is the cosmological constant and we have introduced generic dimensionful parameters, $m^2_{ab}$, $A_{abc}$ and $M_{ij}$, in the scalar and fermion sectors.

This scenario has the remarkable property of renormalizability~\cite{Stelle:1976gc,Barvinsky:2017zlx}. Moreover, for small enough values of $f_2$, namely
\be f_2 \lesssim  10^{-8} \label{smallf2} \ee
the hierarchy problem is solved~\cite{Salvio:2014soa,Salvio:2017qkx}: gravity does not lead to an excessive radiative contribution to the Higgs mass, which can, therefore, obtain naturally~\cite{tHooft:1979rat} its observed value. Indeed, in this case the Higgs field acquires a shift symmetry, which protects its mass from large quantum effects. However, this theory was thought  to suffer from two problems:
\begin{description}
\item[I] ~~~~~The clash between asymptotic freedom (or, more generally, asymptotic safety) and stability (understood as the absence of tachyons)~\cite{Stelle:1977ry,Salvio:2017qkx,Salvio:2018crh,Avramidi:1985ki}: whenever the parameters are chosen to ensure stability, {\it perturbation theory} features a Landau pole in the RG flow of $f_0$. 
\item[II] ~~~~The presence of  a field, with spin-2 and mass 
\be M_2=f_2 \bp/\sqrt{2}, \label{M2Def} \ee
but with an unusual sign of the kinetic term~\cite{Stelle:1976gc} (i.e. a ghost-like field). This is a generic issue of theories with more than two time-derivatives in the Lagrangian, which have a {\it classical} Hamiltonian that is unbounded from below, as shown by Ostrogradsky~\cite{ostro}. The theory above has indeed more than two derivatives because of the Weyl-squared term, $W^2$. 
\end{description}
However, in the following we will review (and extend in the case of Problem {\bf II}), recent works that have shown how to address these issues. The solution of problem {\bf I} is particularly relevant for this work because it leads to the quasi-conformal scenario, as we now describe.

In Ref.~\cite{Salvio:2017qkx} it was pointed out that Problem {\bf I} is an artefact of perturbation theory in $f_0$:  when this parameter grows, perturbativity in $f_0$  is lost, but one can develop an expansion in $1/f_0$  showing that {\it no} Landau poles are present,  provided that all fundamental scalars have asymptotically Weyl-invariant couplings ($1/f_0\to 0$, $\zeta_{ab}\to 0$) and the remaining couplings approach fixed points in the UV limit. In other words, the theory flows to a conformal behavior in the infinite energy limit and, in this way, can be UV complete. Although flowing to conformal gravity at infinite energy is consistent, at  finite energy conformal invariance is broken by the scale anomaly and the $R^2$ term and non-vanishing values of $\zeta_{ab}$ are generated. However, as discussed in~\cite{Salvio:2017qkx},  this is a multiloop effect (see \cite{Hathrell:1981zb,Hathrell:1981gz,Jack:1990eb} and references therein). Therefore, $1/f_0$ and $\zeta_{ab}$ can remain tiny down to the inflationary scales and the subsequent stages of the evolution of the universe. This leads to the quasi-conformal scenario, which is investigated in this work\footnote{Another logical possibility is that the theory features a non-perturbative behavior that drives $f_0$ from a very large value in the UV down to a very small value at the scales of the early universe, which we can access through CMB measurements. This case was studied in~\cite{Kannike:2015apa,Salvio:2017xul}. The purpose of the present paper is instead to explore the quasi-conformal option. }. 

Let us now turn to Problem {\bf II}. Although higher derivatives can potentially be dangerous for a theory, it can be shown that this does not need to be so for the theory we consider here. In order to present a quantitative argument it is necessary to proceed perturbatively: we split $g_{\mu\nu}$ as 
$$  g_{\mu\nu} = g^{\rm cl}_{\mu\nu} + \hat h_{\mu\nu},$$
where $g^{\rm cl}_{\mu\nu}$ is a classical background that solves the classical equations of motion (EOMs)
and $\hat h_{\mu\nu}$ is a  quantum fluctuation. 
Although a non-perturbative approach would be desirable, that is not  currently within the reach of our understanding of quantum field theory\footnote{We want a realistic theory of all forces, but it is not even known how to treat non-perturbatively e.g. chiral fermions (the lattice is currently unable to study fermions with chiral gauge interactions).}. At this point it is clear that the $W^2$ term poses two problems: one that concerns the classical evolution (related to $g^{\rm cl}_{\mu\nu}$) and another one that regards the quantum dynamics (encoded in $\hat h_{\mu\nu}$).

The fundamental question at the classical level is the following: can we avoid the possible instabilities due to the Ostrogradsky theorem? One obvious way to avoid runaways is to take $f_2\gg 1$, such that the Weyl-squared term would have a negligible effect on any experimentally observable quantities. However, one would like to keep $f_2$ small both to see the effects of this term and to solve the hierarchy problem (see~(\ref{smallf2})). Moreover, when the massive spin-2 field is not tachyonic, $M_2^2>0$, it is also asymptotically free: $f_2$ goes to zero in the infinite energy limit~\cite{Fradkin:1981iu,Fradkin:1981hx,Avramidi:1985ki}. In Ref.~\cite{Salvio:2019ewf} it was shown that the answer to the question above is  positive for the theory considered here even for a small $f_2$ (see also Ref.~\cite{Chapiro:2019wua} for a related calculation). The key properties used in~\cite{Salvio:2019ewf} was the fact that the massive spin-2 field decouples as $f_2\to 0$ and is not tachyonic. We here highlight the main steps of the argument in~\cite{Salvio:2019ewf}. 
  \begin{itemize}
\item First, one observes that in the  free-field limit the Hamiltonian of the massive spin-2 field is
    $$H_{\rm 2} \,\, ~ =  \,\, ~ - \sum_{\alpha=\pm 2, \pm 1, 0} \int d^3 q \left[ P_\alpha^2 + (q^2+M_2^2) Q_\alpha^2 \right] $$
    where $Q_\alpha$ and $P_\alpha$ are the associated  canonical variables and conjugate momenta and the helicity sum is over $\alpha= \pm 2, \pm 1, 0$ because this massive particle has spin 2. The manifestation of the Ostrogradsky theorem in this case is the overall minus sign. 
    However, despite that sign there are no instabilities in the free-field limit (that sign cancels in the EOM).
    \item The effective field theory (EFT) approach tells us that at energies below $M_2$ we should not find runaways even if the massive spin-2 field has an order-one coupling, $f_2 \sim 1$.
       \item The intermediate case $0 < f_2 < 1$ must have intermediate energy thresholds
       (above which the runaways might be activated).
        \item   The weak coupling case $f_2 \ll 1$ (compatible with Higgs naturalness)  must have an energy thresholds
        much larger than $M_2$: we could see the effect of the massive spin-2 field without runaways.
    \end{itemize}

The complete argument of~Ref.~\cite{Salvio:2019ewf} leads to the following energy thresholds. When the energies $E$ associated with the derivatives of the spin-2 fields (both the massless graviton and the massive one) satisfy
 \be E\ll  E_2 \equiv \frac{M_2}{\sqrt{f_2}} = \sqrt{\frac{f_2}{2}} \bp  \qquad \mbox{(for the derivatives of the spin-2 fields)}\label{E2} \ee
 and the energies $E$ associated with the matter sector (due to derivatives of matter fields or matter-field values times coupling constants) fulfill
\be E\ll E_m\equiv \sqrt[4]{f_2} \bp \qquad \mbox{(for the matter sector)}\label{Em} \ee
the runaways are avoided. Even for $f_2$ as small as $10^{-8}$ this threshold is high enough to accommodate the whole cosmology\footnote{Furthermore, the fatal runaways above such threshold give an (anthropic) rationale for a homogeneous and isotropic universe~\cite{Salvio:2019ewf}.}~\cite{Salvio:2019ewf}. We will keep in mind these bounds when analysing the quasi-conformal models of the early universe in Secs.~\ref{model} and~\ref{Natural inflation}.

Although the argument above gives a satisfactory solution of the classical Ostrogradsky problem it is still needed to address the issues raised by the quantization. One reason is the fact that quantum effects might lead to tunneling above the energy threshold even if the classical fields satisfy the bounds in~(\ref{E2}) and~(\ref{Em}). However, we know that renormalizability implies that the quantum Hamiltonian governing $\hat h_{\mu\nu}$ is bounded from below~\cite{Stelle:1976gc,Salvio:2018crh} and, therefore, this dangerous tunneling should not occur. On the other hand, renormalizability also implies that the space of states is endowed with an  indefinite metric (with respect to which the quantum mechanical ``position" $q$ and momentum $p$ operators are self-adjoint)~\cite{Stelle:1976gc,Salvio:2015gsi,Salvio:2018crh,Raidal:2016wop} .  The presence of an indefinite metric, therefore, leads to the question: how can we define probabilities consistently?

The crucial observation here is that such metric is not the one  that should be used in the Born rule to compute probabilities. To find the correct norm we start by a definition of observables that generalizes the one usually given in quantum mechanics to avoid any reference to a specific norm. We define ``observable" any operator $A$ with a complete set of eigenstates\footnote{One can show that the basic operators $q$ and $p$ as well as the Hamiltonian in the theory considered here have complete eigenstates at {\it any} order in perturbation theory.}
$\{|a\rangle \}$~\cite{Salvio:2018crh}:
  for any state $|\psi \rangle$ there is a decomposition $$|\psi \rangle = \sum_a c_a |a\rangle$$ 
  for some coefficients $c_a$. 
Moreover, we interpret  $|a\rangle$ as the state where $A$ assumes certainly the value corresponding to $a$ (the deterministic part of the Born rule). 
  The correct way to approach this problem is to recall how experiments are performed. Experimentalists prepare a large number  of times $N$ the same state, so one is naturally  led to consider the direct product of $ |\psi\rangle$ with itself $N$ times:
  $$ |\Psi_N\rangle \equiv  \nu^N|\psi\rangle ... |\psi\rangle = \sum_{a_1 ... a_N} d_{a_1} ...  \, d_{a_N} |a_1\rangle ... |a_N\rangle,$$ 
  where we have introduced a normalization factor $\nu\equiv 1/\sqrt{\sum_b|c_b|^2}$ and correspondingly defined the normalized coefficients $d_{a}\equiv \nu c_a$.
The next step is to define a frequency operator $F_a$ that counts the number  of  times $N_a$ there is the value $a$ in the state $|a_1\rangle ... |a_N\rangle$:  
$$ F_a |a_1\rangle ... |a_N\rangle = \frac{N_a}{N}  |a_1\rangle ... |a_N\rangle.  $$ 
Refs.~\cite{Farhi:1989pm,Strumia:2017dvt} showed that 
 $$ \lim_{N\to\infty} F_a   |\Psi_N\rangle  =  p_a  |\Psi_N\rangle, \qquad p
 _a \equiv \frac{|c_a|^2}{\sum_b|c_b|^2}, $$ 
in the sense that all coefficients in the basis $|a_1\rangle ... |a_N\rangle$ of both $F_a   |\Psi_N\rangle$ and $p_a  |\Psi_N\rangle$ converge to the same quantities. This gives us a strong physical motivation for assuming the complete Born rule to compute probabilities: the probability of observing  $|a\rangle$ if the system is prepared in the state $|\psi \rangle$ is given by $p_a$. The born rule can be expressed in terms of a norm on the space of states, {\it but this is not the indefinite metric mentioned above}:
 \be  p_a \equiv \frac{|c_a^2|}{\sum_b|c^2_b|}  = \frac{|\langle a|\psi\rangle_A|^2}{ \langle \psi|\psi\rangle_A}.\label{Ba}\ee
Indeed, for any pair of states $|\psi_1\rangle$, $|\psi_2\rangle$, we denote  the indefinite metric mentioned above with $\langle \psi_2|\psi_1\rangle$ and the positive norm which gives the Born rule in~(\ref{Ba}) is different and  defined by
\be\langle \psi_2|\psi_1\rangle_A \equiv \langle \psi_2|P_A|\psi_1\rangle_A, \qquad \langle a'|P_A|a\rangle \equiv \delta_{aa'}.\ee 
The second equation above gives the definition of the ``norm" operator $P_A$.
From the definition of the probabilities $p_a$ in Eq.~(\ref{Ba}) it is easy to show that all probabilities are positive 
and that they sum up to one at any time. Therefore, the theory is unitary\footnote{A different proposal for unitarizing this theory was found in~\cite{Anselmi:2017ygm,Anselmi:2019rxg}. In this work we do not adopt this alternative approach because it is not (currently) developed enough to provide us with predictions for the CMB observables. On the other hand, for the approach discussed here, the predictions for such observables have been found in~\cite{Salvio:2017xul}.}.

A remaining issue that has been pointed out is that higher-derivative theories may generate a  violation of micro causality in its decay processes~\cite{Grinstein:2008bg}.  However,  whenever $M_2< H$, where $H$ is the Hubble rate during inflation\footnote{Note that $H$ satisfies the observational bound  $H< 2.7\cdot 10^{-5}\bp \,  (95\%~ \mbox{CL})$~\cite{Ade:2015lrj,Akrami:2018odb}  so $H\ll \bp$ and $M_2< H$ implies $f_2\ll 1$ (see Eq.~(\ref{M2Def})).}, the width of the massive spin-2 field, $\Gamma_2$, (the only one that could lead to micro acausality) is always much smaller than $H$ as $\Gamma_2 \ll M_2$ in this case~\cite{Salvio:2016vxi,Anselmi:2018tmf,Salvio:2018kwh}. Therefore, these potential acausal processes are actually quickly diluted by the expansion of the universe. In the opposite case,  $M_2> H$, the Weyl-squared term does not contribute to the experimentally  observable quantities as it essentially decouples~\cite{Salvio:2017xul}. Therefore, the Weyl-squared term effectively respects  the causality principle

Having addressed the potential issues raised by the $W^2$, we now turn to the analysis of the quasi-conformal scenario.

\section{Hilltop-inflation realization}\label{model}

We describe here a simple realisation of this scenario. The inflaton-gravity (IG) sector features a single scalar field $\phi$ non-minimally coupled to gravity with Lagrangian 
\begin{equation} \Lag_{\rm IG}= \sqrt{-g}\left[\frac{\left(\partial\phi\right)^2}{2} - V(\phi)-\frac{\bp^2 + \xi \phi^2}{2}R -\frac{W^2}{2\gt^2} \right], \label{Jordan-frame-total} \end{equation}
where the non-minimal coupling is taken close to the conformal value $\xi\approx -1/6$ (or $\zeta=\xi+1/6\approx 0$).  
The potential will be  chosen of the form 
 \be V(\phi)=\frac{\lambda_\phi}{4} (\phi^2-v^2)^2. \label{JordanV}\ee 
 We neglect the present-day cosmological constant $\Lambda$ here as completely negligible compared to the early-universe scales.
  Such a potential is the simplest one that guarantees the presence of a maximum at a moderate field value (zero in this case), which in turn can lead to hilltop inflation \cite{Boubekeur:2005zm,Ballesteros:2015iua}, and a minimum where $\phi$ can lie after inflation. This is what one needs to have inflation for $\xi\approx -1/6$, indeed, in order for the effective Planck mass $\sqrt{\bp^2+\xi \phi^2} \approx \sqrt{\bp^2-\phi^2/6}$ to remain real, inflation should occur at moderate field values (not exceeding $\sqrt{6}\bp$).

The non-minimal coupling $-\xi \phi^2 R/2$ can be eliminated through the {\it conformal} transformation
\begin{equation} g_{\mu \nu}\rightarrow   \Omega^{-2}  g_{\mu \nu}, \quad \Omega^2= 1+\frac{\xi \phi^2}{\bp^2}. \label{transformation}\end{equation}
Note that, in order for this field redefinition to be well-defined, we have to require $|\phi|<\bp/\sqrt{|\xi|}$ when $\xi<0$, so in particular this constraint applies for $\xi = -1/6$. This is nothing but the requirement of having a real effective Planck mass discussed above. In the  frame obtained with the transformation above (called the Einstein frame) gravity is canonically normalized.
Indeed, the action (after the conformal transformation) becomes 
\begin{equation}  \Lag_{\rm IG}= \sqrt{-g}\left[K(\phi) \frac{(\partial \phi)^2}{2}-\frac{V}{\Omega^4}-\frac{\bp^2 }{2}R -\frac{W^2}{2\gt^2} \right], \end{equation}
where $$K(\phi) \equiv \frac1{\Omega^2}+\frac{6\xi^2\phi^2/\bp^2}{\Omega^4} ~~~~\stackrel{\mathclap{\normalfont \xi\approx -1/6}}{\approx} ~~~~ \frac{1}{\left[1-\phi^2/(6\bp^2)\right]^2}.
$$

The non-canonical kinetic term can be made canonical through the field redefinition $\phi=\phi(\chi)$ defined by
\begin{equation} \frac{d\chi}{d\phi}= \sqrt{K(\phi)},\label{chi}\end{equation}
with the conventional condition\footnote{For a generic number $N_s$ of scalar fields $\phi_a$ ($a=1..., N_s$) Eq.~(\ref{Jordan-frame-total}) gets replaced by
\begin{equation} \Lag_{\rm IG}= \sqrt{-g}\left[\frac{\left(\partial\phi_a\right)^2}{2} - V(\phi)-\frac{\bp^2 + \xi_{ab} \phi_a\phi_b}{2}R -\frac{W^2}{2\gt^2} \right],\label{footnote1} \end{equation} 
and going to the Einstein frame one finds
\begin{equation}  \Lag_{\rm IG}= \sqrt{-g}\left[K_{ab}(\phi) \frac{\partial \phi_a\partial \phi_b}{2}-\frac{V}{\Omega^4}-\frac{\bp^2 }{2}R-\frac{W^2}{2\gt^2} \right],\label{footnote2} \end{equation}
where
\be \Omega^2= 1+\frac{\xi_{ab} \phi_a\phi_b}{\bp^2}\qquad \stackrel{\mathclap{\normalfont \xi_{ab}\approx -\delta_{ab}/6}}{\approx}\qquad 1-\frac{\phi_a\phi_a}{6\bp^2}, \label{footnote2b} \ee
and 
\be  K_{ab}=  \frac{\delta_{ab}}{\Omega^2} +\frac{3\bp^2\partial_{a}\Omega^2\partial_{b}\Omega^2}{2\Omega^4}=  \frac{\delta_{ab}}{\Omega^2} +\frac{6\xi_{ca}\phi_c\xi_{db}\phi_d}{\bp^2 \Omega^4}\qquad \stackrel{\mathclap{\normalfont \xi_{ab}\approx -\delta_{ab}/6}}{\approx}\qquad \frac{\delta_{ab}\left[1-\phi_c\phi_c/(6\bp^2)\right]+\phi_a\phi_b/(6\bp^2)}{\left(1-\phi_c\phi_c/(6\bp^2)\right)^2}.  \label{footnote3} \ee
 Generically, the Ricci scalar of the field metric $K_{ab}$ is not zero  so  $K_{ab}$ is not flat.} $\phi(\chi=0)=0$. One can find a closed expression of $\chi$ as a function of $\phi$ \cite{Salvio:2015kka}:
\be \chi(\phi) =  \bp \sqrt{\frac{1+6 \xi }{\xi}} \,\text{arcsinh}\left[\frac{\sqrt{\xi  (1+6 \xi )}\phi }{\bp}\right]-\sqrt{6} \bp \,\text{arctanh}\left[\frac{\sqrt{6} \xi \phi }{\sqrt{\bp^2+\xi  (1+6 \xi )\phi ^2}}\right].\ee
The function $\chi(\phi)$ can be inverted to obtain $\phi(\chi)$ because Eq.~(\ref{chi}) implies $d\chi/d\phi > 0$. 
Thus, $\chi$ feels a potential 
\begin{equation} V_E(\chi)\equiv \frac{V(\phi(\chi))}{\Omega^4(\phi(\chi))}=\frac{\lambda_\phi(\phi(\chi)^2-v^2)^2}{4(1+\xi\phi(\chi)^2/\bp^2)^2}\label{VE} .\end{equation}
The label $E$ recalls that this is the Einstein frame potential, as opposed to the so-called Jordan frame one, Eq.~(\ref{JordanV}).
It is difficult to compute analytically $\phi(\chi)$ for a generic value of $\xi$. But for $\xi = -1/6$ we have simply 
\be \chi(\phi) = \sqrt{6} \bp \, \text{arctanh}\left(\frac{\phi}{\sqrt{6 }\bp}\right) \ee
and therefore 
\be \phi(\chi) = \sqrt{6} \bp \, \text{tanh}\left(\frac{\chi}{\sqrt{6 }\bp}\right). \label{phiOFchi}  \ee
This is the reason why an inflationary scenario with a quasi-conformally coupled inflaton can be analytically studied. In the following, as an example, we will present the analytic inflationary formul\ae\, in the case of hilltop inflation, with Jordan frame potential~(\ref{JordanV}), but an analytic treatment can always be carried out whenever one has a closed form for the Jordan frame potential. 

By inserting~(\ref{phiOFchi}) inside~(\ref{VE}) one finds
\be V_E(\chi)  =   \frac{\lambda_\phi}{4}   \, \text{cosh}^4\left(\frac{\chi}{\sqrt{6} \bp}\right) \left(v^2-6 \bp^2 \tanh ^2\left(\frac{\chi }{\sqrt{6} \bp}\right)\right)^2. \label{VEofchi}\ee
Note that this potential does not have a singularity for any value of $\chi$ although $V(\phi)/\Omega^4(\phi)$ diverges at $\phi = \sqrt{6} \bp$ (having set $\xi = -1/6$). The reason why this happens is because the function $\phi(\chi)$ in~(\ref{phiOFchi})   reaches the value $\sqrt{6} \bp$ only for  $\chi \to \infty$, which  is due to the fact that the singularity of $V(\phi)/\Omega^4(\phi)$ at $\phi = \sqrt{6} \bp$ is compensated by a singularity of $K(\phi)$ at the same point. There is an analogous effect for generic values of $\xi < 0 $~\cite{Linde:2011nh}. Also note that $V_E$ has two stationary points for $v< \sqrt{6} \bp$: a maximum at $\chi = 0$ and a vanishing minimum at $\chi = \sqrt{6} \bp\text{arctanh}\left(v/\sqrt{6} \bp\right)$; for $v> \sqrt{6} \bp$ the latter minimum disappears and the only  stationary point is $\chi = 0$, which becomes a minimum. However, as commented before, for $v> \sqrt{6} \bp$ the effective Planck mass $\sqrt{\bp^2 -v^2/6}$ becomes imaginary and the conformal transformation in~(\ref{transformation}) is ill-defined, thus one must exclude this case.

\begin{figure}[t]
  \includegraphics[scale=0.55]{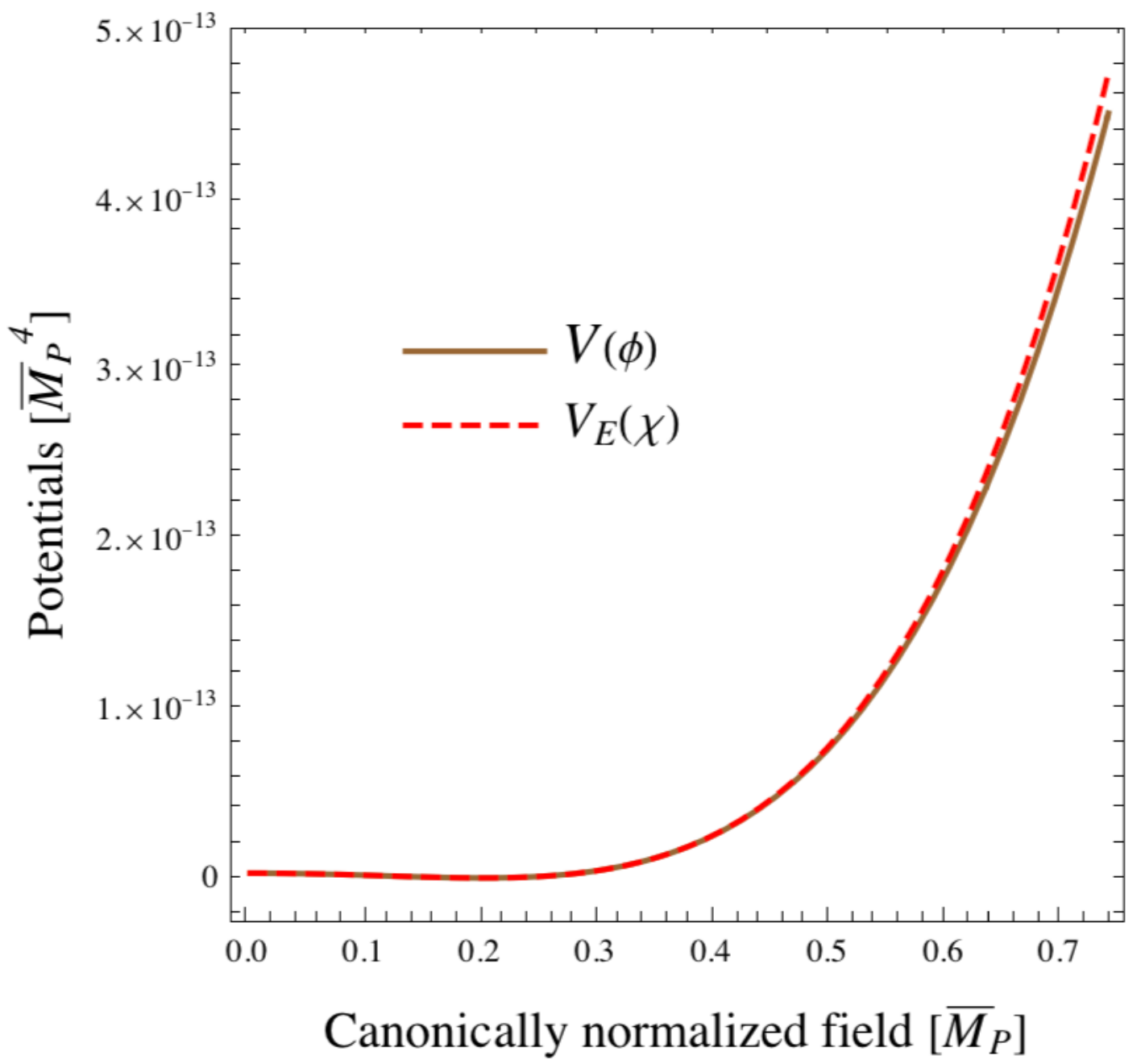}  \hspace{1.1cm} \includegraphics[scale=0.55]{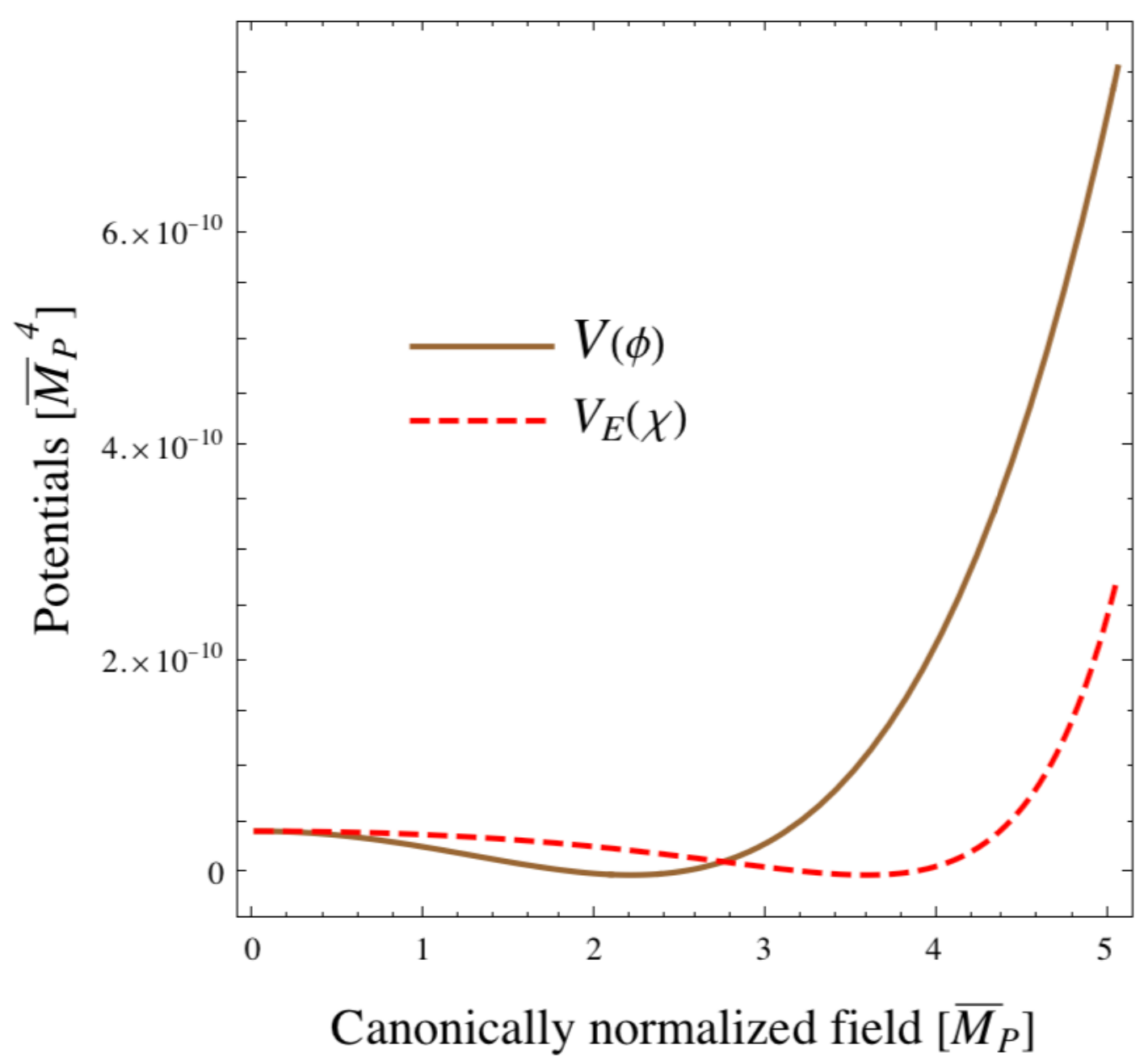}  
   \caption{\em  Jordan frame and Einstein frame potentials (Eqs.~(\ref{JordanV}) and~(\ref{VEofchi})) as a function of the respective canonically normalized fields, $\phi$ and $\chi$. In both plots $\lambda_\phi= 7 \times 10^{-12}$.  On the left we set $v= 0.2 \bp$, while on the right $v = 2.2 \bp$.}
\label{ns-and-r}
\end{figure}

Fig.~\ref{ns-and-r} gives $V$ as a function of $\phi$ and $V_E$ as a function of $\chi$ for two values of $v$.  $V$  and $V_E$ are very similar for small enough field values. However, in the opposite limit they differ.

\subsection{Computing inflationary observables}\label{Computing inflationary observables}

Let us take as a starting point the Friedmann-Robertson-Walker (FRW) metric
 \be ds^2 = dt^2 -a(t)^2 \left[\frac{dr^2}{1- k r^2} +r^2(d\theta^2 +\sin^2\theta \,d\varphi^2)\right],  \label{FRW}\ee
 ($k=0,\pm 1$). 
 Correspondingly, we assume the scalar field $\chi$ to be a function of $t$ only.
 
 Then the Einstein equations  and the scalar equations imply the following equations for $a(t)$ and the spatially homogeneous field $\chi(t)$  
  \bea
  \ddot \chi  + \frac{3\dot a}{a}\dot \chi+V_E'  &=&0,\label{scalarEq} \\  
  \frac{\dot a^2+k}{a^2}-\frac{ \dot\chi^2+2V_E}{6 \bar M_{\rm Pl}^2} &=&0 \label{EE1}, \\ 
  \frac{k}{a^2}-\frac{d}{dt}  \frac{\dot a}{a}-\frac{\dot \chi^2}{ 2\bar M_{\rm Pl}^2}&=&0, \label{EE2}\eea
where a dot denotes a derivative with respect to $t$ and a prime is a derivative with respect to $\chi$.
   From Eqs.~(\ref{scalarEq}) and~(\ref{EE1}) one can derive~(\ref{EE2}), which is, therefore, dependent.

During inflation we assume that the energy density is dominated by the scalar field $\chi$ and set $k=0$;
moreover,  the following slow-roll parameters should be small
\be \epsilon \equiv\frac{\bp^2}{2} \left(\frac{V_E'}{V_E}\right)^2, \quad \eta \equiv \bp^2\frac{V_E''}{V_E} .\label{epsilon-def}\ee
For $\xi=-1/6$ one obtains the analytic expressions
\be \epsilon(\chi) =\frac{4 \left(v^2-6 \bp^2\right)^2 \tanh ^2\left(\frac{\chi}{\sqrt{6} \bp}\right)}{3 \left(v^2-6 \bp^2 \tanh ^2\left(\frac{\chi}{\sqrt{6} \bp}\right)\right)^2},\ee
\be \eta(\chi) = \frac{6 \left(6 \bp^2-v^2\right) \text{sech}^2\left(\frac{\chi}{\sqrt{6} \bp}\right) \left(2 \bp^2 \text{sech}^2\left(\frac{\chi}{\sqrt{6} \bp}\right)-10 \bp^2+v^2\right)+8 \left(v^2-6 \bp^2\right)^2}{3 \left(v^2-6 \bp^2 \tanh ^2\left(\frac{\chi}{\sqrt{6} \bp}\right)\right)^2}. \ee 
In the slow-roll approximation the EOMs~(\ref{scalarEq})-(\ref{EE2}) are 
 \bea
\dot \chi  &=& -\frac{V_E'}{3H},\label{scalarEqp} \\  
 \frac{\dot a^2}{a^2}&=&\frac{V_E}{3 \bar M_{\rm Pl}^2}  \label{EE1p}, \\ 
  \frac{\dot \chi^2}{ 2\bar M_{\rm Pl}^2}&=&-\frac{d}{dt}  \frac{\dot a}{a}. \label{EE2p}\eea
  The scalar spectral index $n_s$ and the tensor-to-scalar ratio $r$ are given by~\cite{Salvio:2017xul}
\bea n_s&=&1-6\epsilon +2\eta, \label{nsformula} \\ r_E&=&16\epsilon, \label{rformula}   \\ r &=& \frac{r_E}{1+\frac{2 H^2}{M_2^2}},\label{rWformula}\eea
where $r_E$ is the tensor-to-scalar ratio when the pure gravitational Lagrangian is the Einstein-Hilbert one~\cite{Salvio:2017xul}. The difference between $r$ and $r_E$ is due to $W^2$. The Weyl-squared term (if  $M_2<H$) also leads to an isocurvature mode with power spectrum $P_B$, such that the ratio 
\be r' \equiv P_{B}/P_R\label{rp}\ee
 is a factor $3/16$ smaller than $r_E$~\cite{Salvio:2017xul}.    Eqs.~(\ref{rWformula}) and~(\ref{rp}) are valid in a generic inflationary model~\cite{Salvio:2017xul} and so they apply to the natural-inflation model that will be studied in Sec.~\ref{Natural inflation} too. A general observational bound on $r'$ have been set in Ref.~\cite{Salvio:2019ewf} (see Fig.~7 in that article), which translates into a bound on $r_E$ because, as mentioned above, $r'=3r_E/16$.

Moreover, we find the following analytic expressions
$$ n_s(\chi) = \frac{12\, \text{sech}^2\left(\frac{\chi}{\sqrt{6} \bp}\right) \left[\bp^2 \left(21 \bp^2-2 v^2\right) \text{sech}^2\left(\frac{\chi}{\sqrt{6} \bp}\right)-6 \bp^4-5 \bp^2 v^2+v^4\right]-5 \left(v^2-6 \bp^2\right)^2}{3 \left(v^2-6 \bp^2 \tanh ^2\left(\frac{\chi}{\sqrt{6} \bp}\right)\right)^2},$$
\be r_E(\chi) = \frac{64 \left(v^2-6 \bp^2\right)^2 \tanh ^2\left(\frac{\chi}{\sqrt{6} \bp}\right)}{3 \left(v^2-6 \bp^2 \tanh ^2\left(\frac{\chi}{\sqrt{6} \bp}\right)\right)^2}.  \ee 
and, since in the slow-roll approximation $H$  is given by $H^2(\phi) = V_E(\phi)/3\bp^2$, we have 
 \be  r(\chi) = \frac{64 \left(v^2-6 \bp^2\right)^2 \tanh ^2\left(\frac{\chi}{\sqrt{6} \bp}\right)}{3 \left(v^2-6 \bp^2 \tanh ^2\left(\frac{\chi}{\sqrt{6} \bp}\right)\right)^2\left[1+2 V_E(\chi )/(3M_2^2\bp^2)\right]}  .\ee
 
  The number of e-folds $N_e$ as a function of the field $\chi$ is given by 
  \be N_e(\chi) = {\cal N}(\chi)  - {\cal N}(\chi_{\rm end})   \label{Ne}   \ee
  where, in the slow-roll approximation, the function ${\cal N}$ is
  \be {\cal N}(\chi)  = \frac1{\bp^2}  \int^\chi d\bar\chi  \,  \frac{V_E(\bar\chi)}{V_E'(\bar\chi)}  \ee 
  and $\chi_{\rm end}$ is the field value at the end of inflation, determined by $\epsilon(\chi_{\rm end}) = 1$: we find an analytic expression for $\chi_{\rm end}$ too,
\be \chi_{\rm end} =   \sqrt{6} \bp \, \text{arctanh}\left(\sqrt{\frac{36 \bp^4-3 \bp^2 v^2+v^4-\sqrt{\left(v^2-6 \bp^2\right)^2 \left(36 \bp^4+6 \bp^2 v^2\right)}}{54 \bp^4}}\right). \ee 
The integration constant in the expression of ${\cal N}(\chi)$ above plays no role because what matters is the difference between ${\cal N}$ computed in different points, Eq.~(\ref{Ne}). Also, one obtains a closed form for  ${\cal N}(\chi)$: 
    \be  {\cal N}(\chi)=\frac{18 \bp ^2 \log \left(\cosh \left(\frac{\chi}{\sqrt{6} M}\right)\right)-3 v^2 \log \left(\sinh \left(\frac{\chi}{\sqrt{6} M}\right)\right)}{12 \bp^2-2 v^2} + \text{const} ,  \ee 
    where ``const" represents an arbitrary constant with respect to $\chi$.

    By requiring that the measured power spectrum \cite{Akrami:2018odb},
\begin{equation}P_{R}= \frac{V_E/ \epsilon}{24\pi^2 \bp^4}= (2.10 \pm 0.03) \times 10^{-9} , \label{normalization} \end{equation}
is reproduced for a field value corresponding to an appropriate value of e-folds one fixes $\lambda_\phi$.

\begin{figure}[!ht]
\begin{center}
\hspace{-0.37cm}
  \includegraphics[scale=0.515]{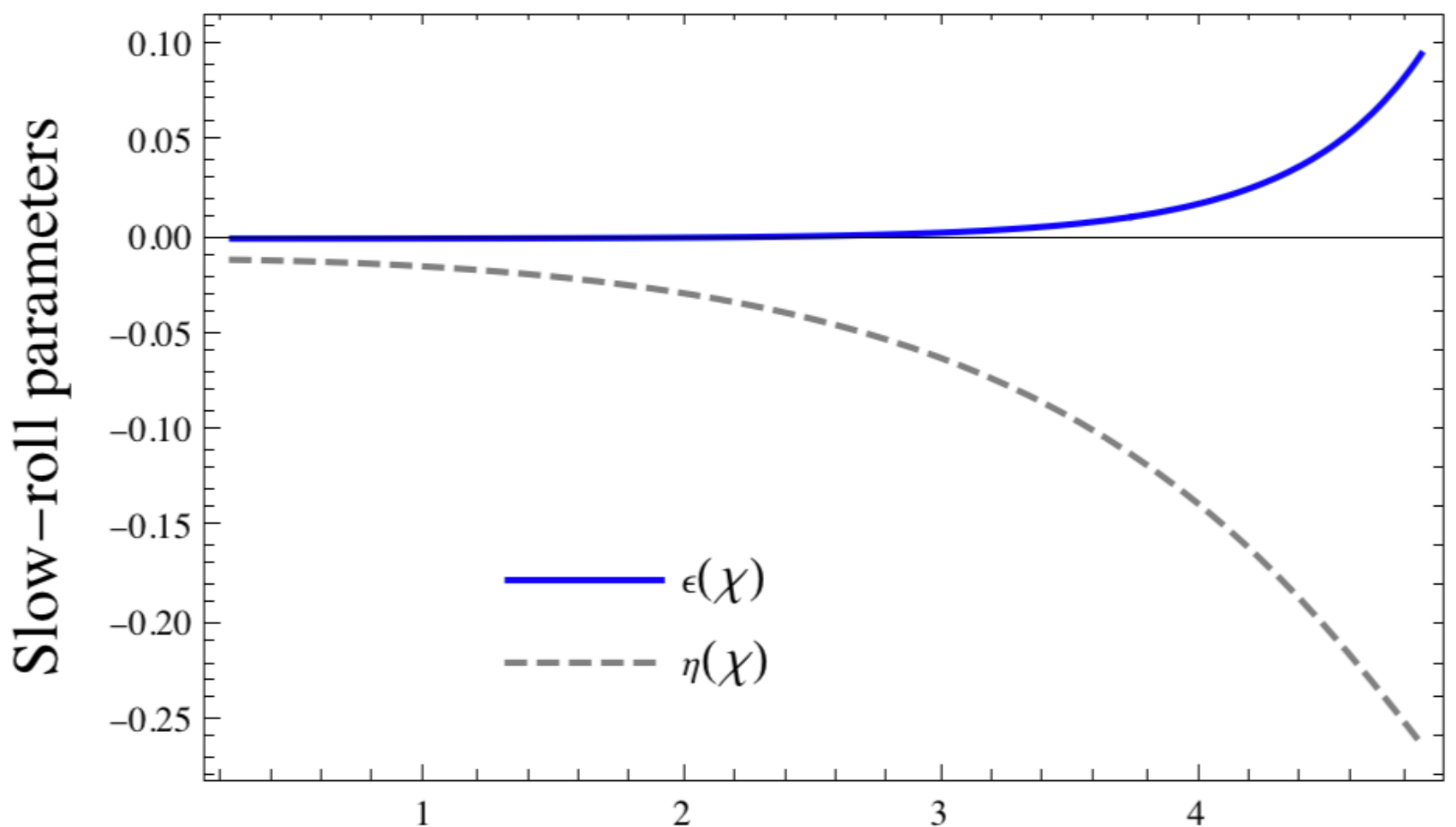} \\
 \includegraphics[scale=0.5]{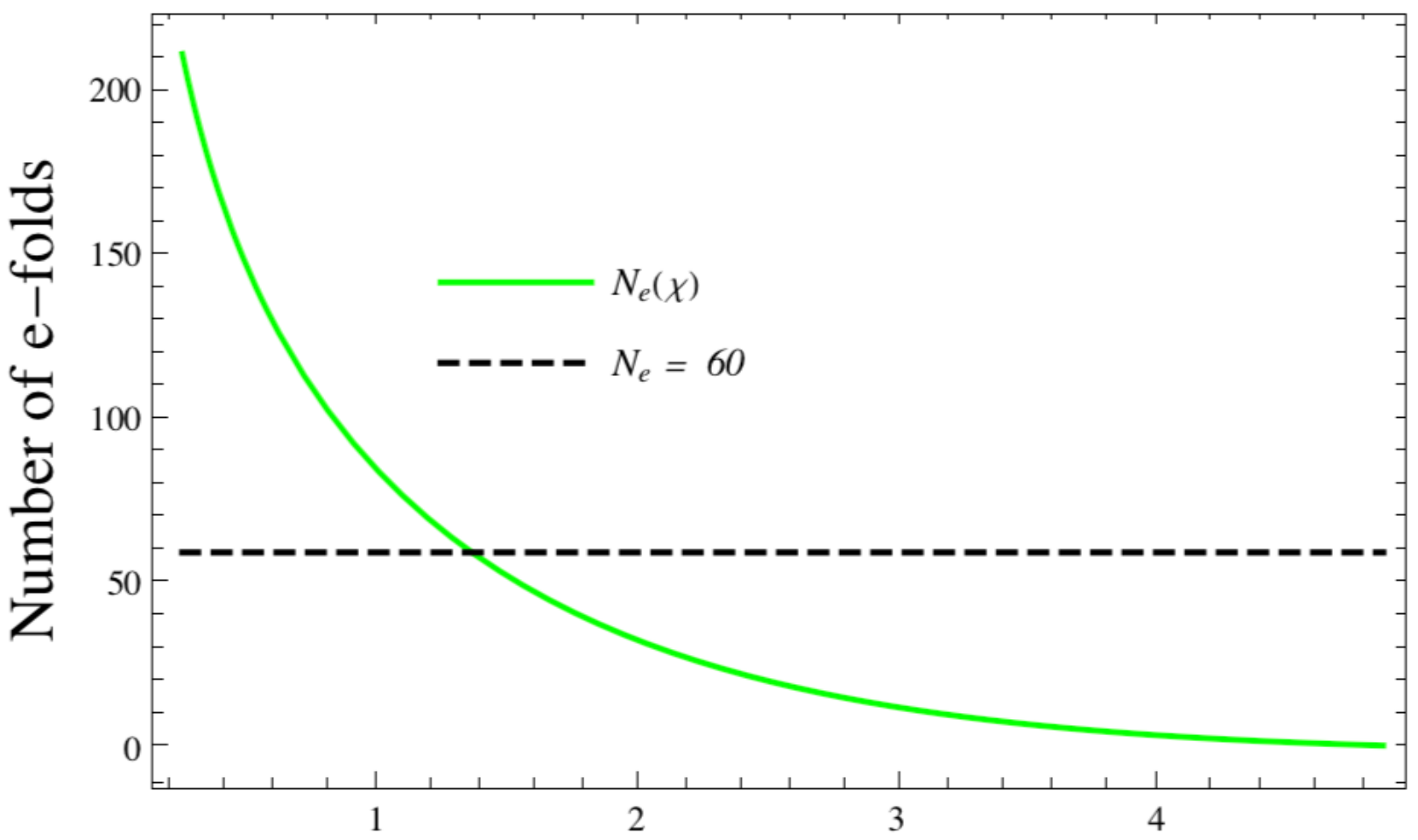}\\ \includegraphics[scale=0.5]{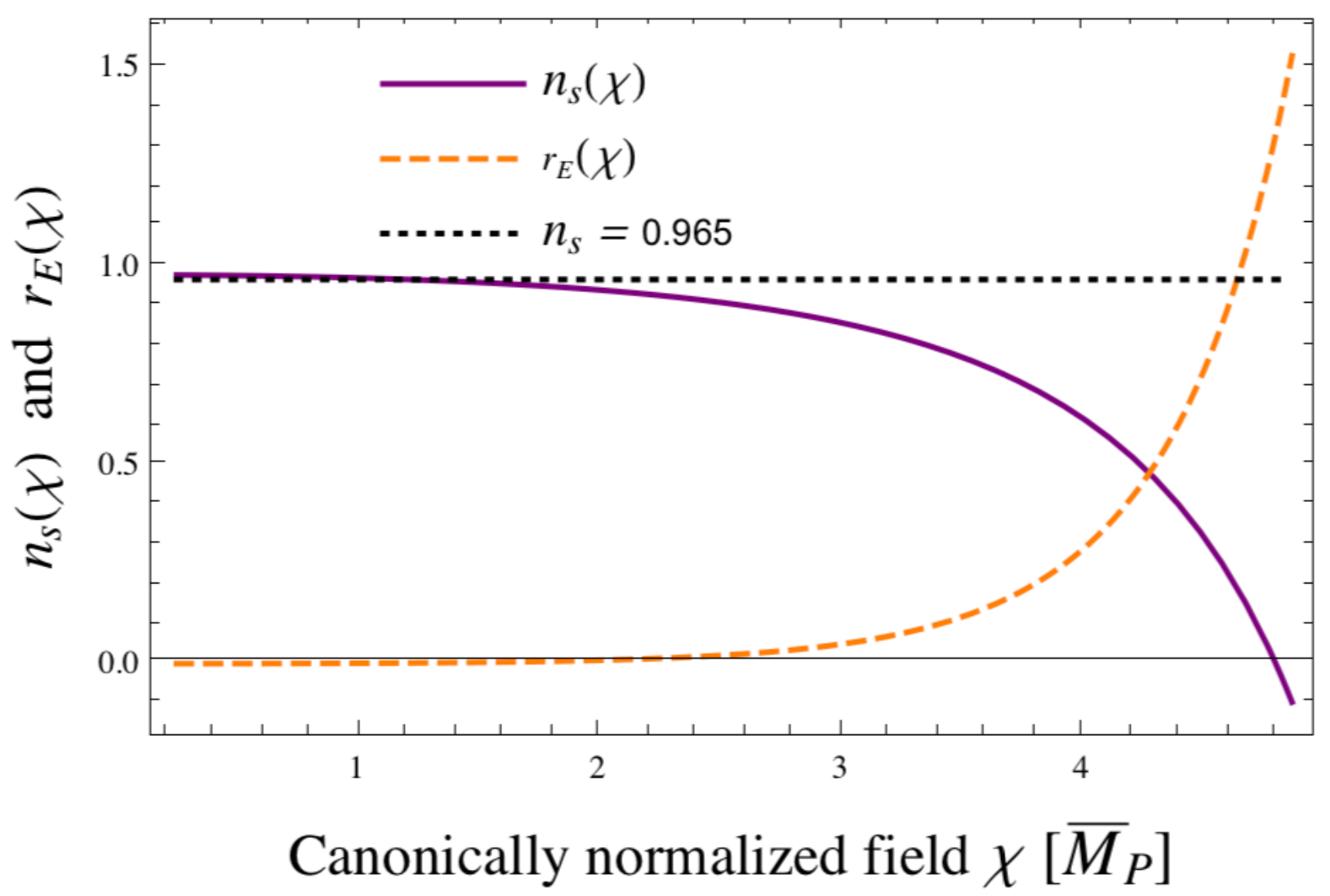} 
 \end{center} 
   \caption{\em The slow-roll parameters $\epsilon$ and $\eta$ defined in Eq.~(\ref{epsilon-def}), the number of e-folds $N_e$ given in~(\ref{Ne}) and $n_s$ and $r$ (computed with Eqs.~(\ref{nsformula}) and~(\ref{rformula})) as a function of $\chi$.  We have set $v = 2.43 \bp$ (the quantities plotted do not depend on $\lambda_\phi$).}
\label{inflation}
\end{figure}

In Fig.~\ref{inflation} it is shown that there are initial field values $\chi$ for which we can have $N_e \sim 60$ e-folds and obtain $n_s$ and $r_E$ (and, therefore, $r$) in good agreement with the bounds in \cite{Ade:2015lrj,Akrami:2018odb}, preserving the slow-roll conditions, $\epsilon\ll 1$, $\eta\ll 1$. Since $r_E< 0.1$~\cite{Salvio:2019ewf} the isocurvature power spectrum parameterized by $r'$ also satisfies the bounds in \cite{Ade:2015lrj,Akrami:2018odb}. By choosing $\lambda_\phi \approx 7 \times 10^{-12}$ one obtains the measured $P_R$ in Eq.~(\ref{normalization}).
A numerical analysis shows that, generically, one has a good agreement with the inflationary observables if $v$ is close enough to $\sqrt{6} \bp$. 

Finally, we have explicitly checked in this hilltop implementation of the quasi-conformal scenario that the bounds in~(\ref{E2}) and~(\ref{Em}) are satisfied during the whole cosmology and, therefore, the possible runaways due to the Ostrogradsky theorem are avoided. 
 
\subsection{Higgs naturalness}\label{Naturalness}

In the previous section we have seen that successful inflation can occur if the vacuum expectation value (VEV) $v$ in the inflaton sector is around (but not exceeding much) the Planck scale.

 The fluctuation of $\phi$ around $v$ acquires a mass $ M_\phi  \equiv \sqrt{2 \lambda_\phi} \, v,$ 
which, for the typical orders of magnitude $v \sim \bp$ and $\lambda_\phi \sim 10^{-11}$ (obtained from the amplitude of the curvature fluctuations) is of order $10^{13}$ GeV. Given that $M_\phi$ lead to a quantum correction to the Higgs mass squared
\be \delta M^2_h \sim \frac{\lambda_{h\phi} M_\phi^2}{(4\pi)^2}, \ee
where $\lambda_{h\phi}$ is the quartic portal coupling between $\phi$ and the Higgs (appearing in the potential as $+\lambda_{h\phi} \phi^2 h^2$, where $h$ is the physical Higgs field),
one obtains an upper bound on $\lambda_{h\phi}$ to ensure the naturalness of the Higgs mass, which requires $\delta M_h^2 \lesssim M^2_h$~\cite{Farina:2013mla}. For the orders of magnitude mentioned above we obtain 
\be \lambda_{h\phi} \lesssim 10^{-20}. \label{NatInf} \ee This is a tiny number, but one should bear in mind that $\lambda_{h\phi}$ is the only coupling between the inflaton and the observable sector (besides gravity) and, therefore, a small $\lambda_{h\phi}$ is not destabilized by quantum corrections. Gravity can take part in the running of $\lambda_{h\phi}$, but the agravity scenario studied in~\cite{Salvio:2014soa} softens gravity above a certain energy scale and its contribution turns out to be negligibly small when the condition in~(\ref{smallf2}) is used. The conditions in~(\ref{smallf2}) and~(\ref{NatInf}) ensure that the shift symmetry of the Higgs field is good enough to protect its mass from gravity and inflaton quantum corrections, respectively, leading to a technically natural~\cite{tHooft:1979rat} Higgs mass.

\subsection{Complexity of UV-completions}

The hilltop inflationary model provides us with an existing proof of quasi-conformal models of inflation. Given that such scenario is motivated by possible UV-completions of Einstein gravity one eventually want to embed hilltop inflation in a  field theory without a high-momentum cutoff. In Refs.~\cite{Giudice:2014tma,Holdom:2014hla,Pelaggi:2015kna} it was shown that renormalizable field theories can be made asymptotically free with respect to all couplings (including the quartic couplings $\lambda_{abcd}$) if embedded in a  theory with a  (semi-)simple gauge group and with a complex field content. The field content can be made simpler in asymptotically safe extensions of the SM, which, however, present some challenges~\cite{Pelaggi:2017abg,Alanne:2019vuk}. The scenario described in Sec.~\ref{The quasi-conformal scenario} is general enough to embed this SM extensions.

Such an embedding for the specific case of hilltop inflation, although interesting, goes beyond the scope of the present work because, as we will see in Sec.~(\ref{Natural inflation}), there are simple and elegant asymptotically free theories in which the inflaton is identified with a composite scalar. In these theories all fundamental scalars feature a quasi-conformal non-minimal coupling, but, as will be discussed in Sec.~\ref{Natural inflation}, the inflaton, being a pseudo-Goldstone boson, has a negligible (and thus non-conformal) non-minimal coupling.

\section{Natural-inflation realizations }\label{Natural inflation}

Another way of implementing the quasi-conformal scenario for the early universe is  identifying the inflaton $\phi$ with the pseudo-Goldstone boson associated with the breaking of a global symmetry. In the context of the non-conformal Einstein-gravity case, this scenario has been proposed in~\cite{Freese:1990rb} and is known as ``natural inflation" because the potential of a pseudo-Goldstone boson is naturally flat thanks to Goldstone theorem. This good feature is preserved when one constructs a natural-inflation model in the quasi-conformal scenario, but, as we will see, some predictions differ with respect to the Einstein-gravity case, such that one can distinguish between these two proposals.

First note that (pseudo-)Goldstone bosons 
 do not need to have a Weyl-invariant non-minimal coupling to gravity in the (even exact) conformal scenario; their $\xi$-couplings can be different from $-1/6$. Indeed, if $\Phi$ is a scalar field in some representation of the spontaneously broken group $G_S$, its non-minimal coupling is proportional to $\xi_{\Phi}|\Phi|^2R$, where $|\Phi|^2$ is invariant under $G_S$, and the Goldstone boson (being morally some phase of $\Phi$) does not feature non-minimal couplings.

The inflaton potential of natural inflation $V_N(\phi)$ can be written as~\cite{Freese:1990rb} 
\be V_N(\phi) = \Lambda^4 \left(1+\cos\left(\frac{N\phi}{f}\right)\right), \label{VInf}\ee
where $\Lambda$ and $f$ are two energy scales and  $N$ is a natural number.
Given that $V_N$ is even  and periodic with period $2\pi f/N$ we restrict ourselves to  the interval\footnote{As far as inflation is concerned, it would be equivalent to consider the  potential with a different sign, $\Lambda^4 \left(1-\cos\left(\frac{N\phi}{f}\right)\right)$, and to restrict ourselves to the interval $\phi \in [\pi f/N, 2\pi f/N]$. Therefore, the two signs (considered in~\cite{Freese:1990rb}) are both covered here.} $\phi \in [0, \pi f/N]$. Like in Sec.~\ref{Computing inflationary observables}, we take an FRW metric with $k=0$ to describe inflation, when the energy density of the universe was dominated by the inflaton contribution.

The slow-roll quantities
\be \epsilon \equiv\frac{\bp^2}{2} \left(\frac{V_N'}{V_N}\right)^2, \quad \eta \equiv \bp^2\frac{V_N''}{V_N} .\label{epsilon-eta-def}\ee
are given by
\be \epsilon =\frac{\bp^2 N^2 \tan ^2\left(\frac{N \phi }{2 f}\right)}{2 f^2}, \quad \eta =-\frac{\bp^2 N^2 \cos \left(\frac{N \phi }{f}\right)}{f^2 \left(1+\cos \left(\frac{N \phi }{f}\right)\right)} .\label{epsilon-eta}\ee

In the slow-roll approximation the spectral index $n_s$ and the tensor-to-scalar ratio in Einstein gravity are given by
\be n_s = 1- 6\epsilon +2\eta, \quad r_E =16\epsilon \label{ns-r}\ee
So
\be n_s =1+\frac{\bp^2 N^2}{f^2}-\frac{2 \bp^2 N^2}{f^2} \sec ^2\left(\frac{N \phi }{2 f}\right), \quad r_E =\frac{8 \bp^2 N^2}{f^2} \tan ^2\left(\frac{N \phi }{2 f}\right). \label{ns-rN}\ee
As discussed before, adding the Weyl-squared term leaves $n_s$  unchanged, but gives the following tensor-to-scalar ratio~\cite{Salvio:2017xul}
\be r =\frac{r_E}{1+2 H^2/M_2^2}, \label{rGen}\ee
where $H$ is the inflationary Hubble rate, which, in the slow-roll approximation, is given by $H^2(\phi) = V_N(\phi)/3\bp^2$. Therefore, for natural inflation
\be r = \frac{8 \bp^2 N^2 \tan ^2\left(\frac{N \phi }{2 f}\right)}{f^2 \left[1+2 V_N(\phi )/(3M_2^2\bp^2)\right]}. \ee 
Note that when $H\gg M_2$ the tensor-to-scalar ratio is much smaller than the one predicted within Einstein gravity, as clear from~(\ref{rGen}). This is interesting because natural inflation is on the verge of being excluded when implemented in Einstein gravity~\cite{Akrami:2018odb}. Therefore, the Weyl-squared term can rescue this well-motivated model when $H> M_2$. 

The number of e-folds $N_e$ as a function of the field $\phi$ is here given by 
  \be N_e(\phi) = {\cal N}(\phi)  - {\cal N}(\phi_{\rm end})    \ee
  where, in the slow-roll approximation, the function ${\cal N}$ is
  \be {\cal N}(\phi)  = \frac1{\bp^2}  \int^\phi d\bar\phi  \,  \frac{V_N(\bar\phi)}{V_N'(\bar\phi)}  \ee 
and $\phi_{\rm end}$ is defined by $\epsilon(\phi_{\rm end}) = 1$, which gives
  \be \phi_{\rm end}=  \frac{2f\arctan\left(\frac{\sqrt{2}f}{N\bp}\right)}{N}.\ee
  One also obtains an analytic expression for ${\cal N}(\phi)$:
  \be  {\cal N}(\phi)=-\frac{2 f^2 \log \left(\sin \left(\frac{N \phi}{2 f}\right)\right)}{\bp^2 N^2} + \text{const} ,  \ee

  \begin{figure}[!t]
\begin{center}
  \includegraphics[scale=0.527]{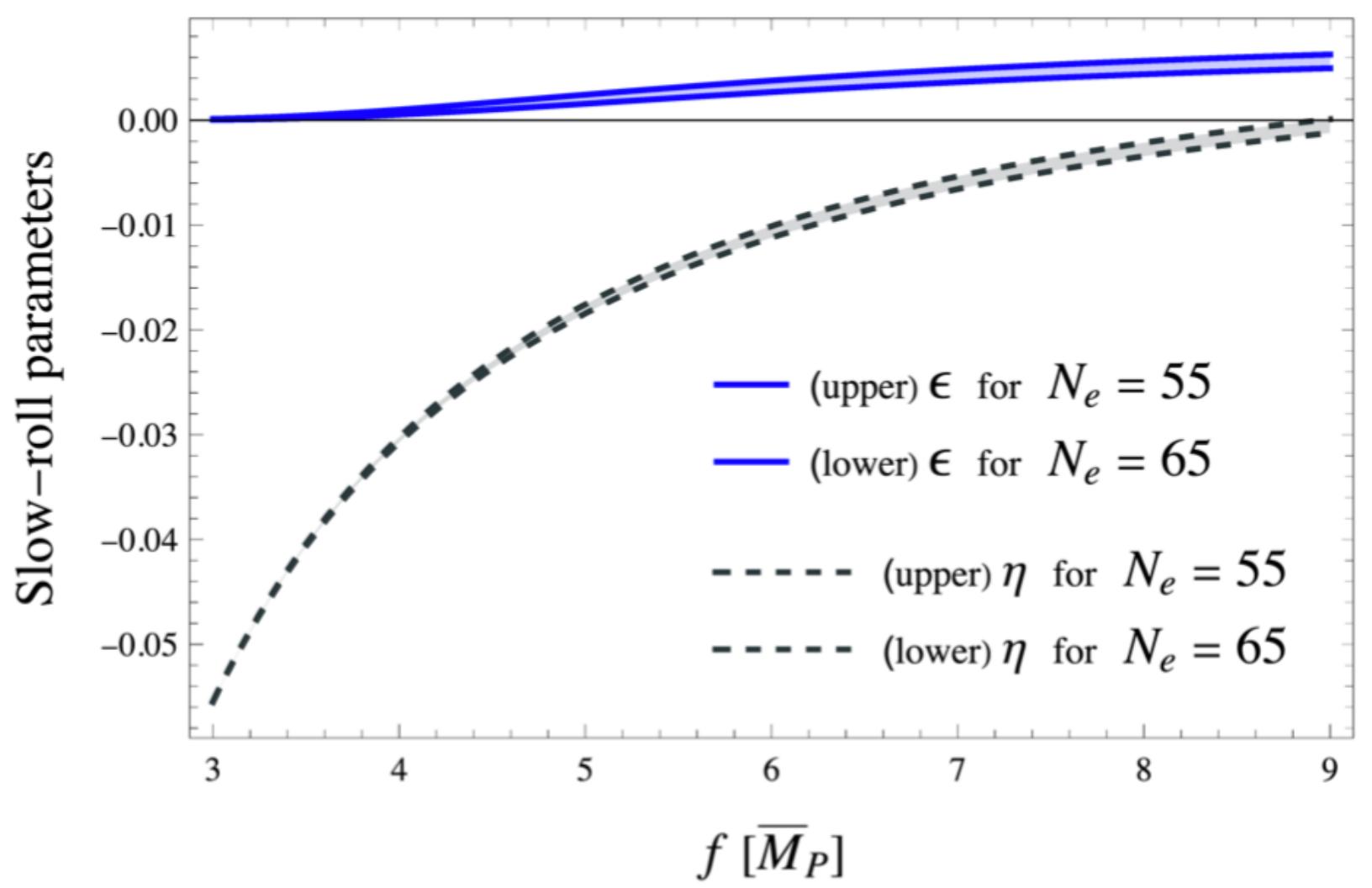} \,\,
   \includegraphics[scale=0.52]{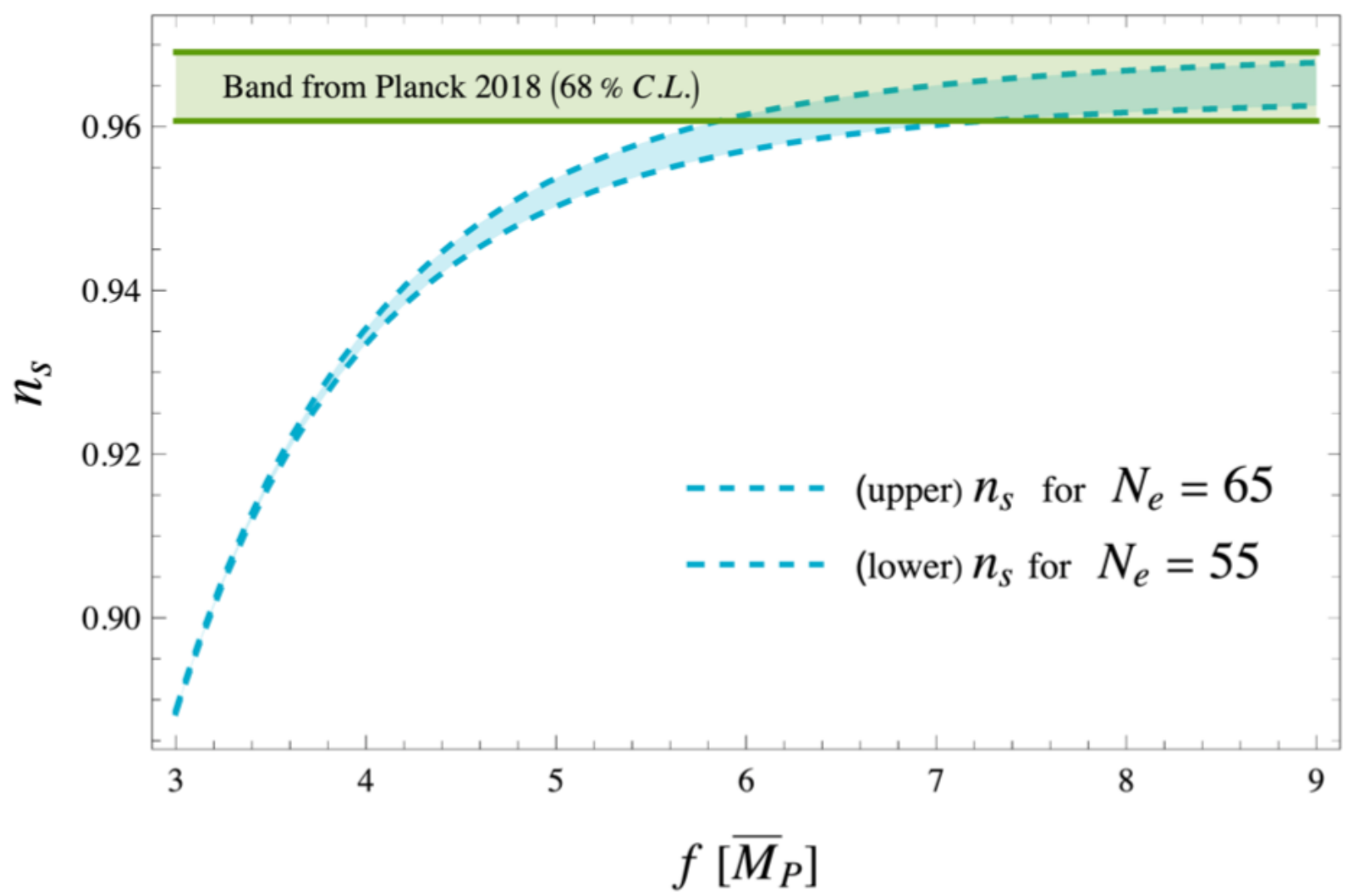} 
   \vspace{0.4cm}
   
     \includegraphics[scale=0.525]{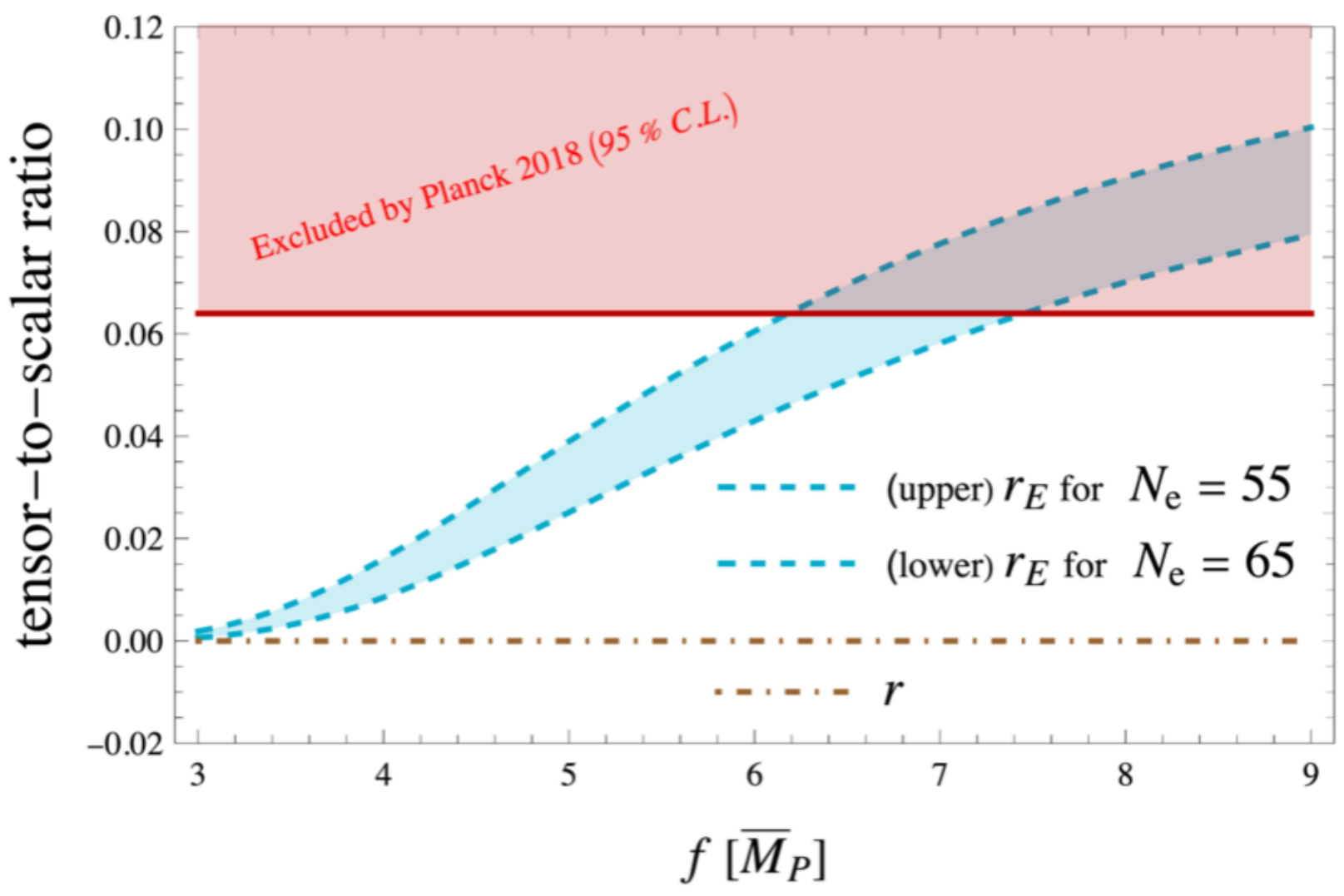}\,
     \includegraphics[scale=0.544]{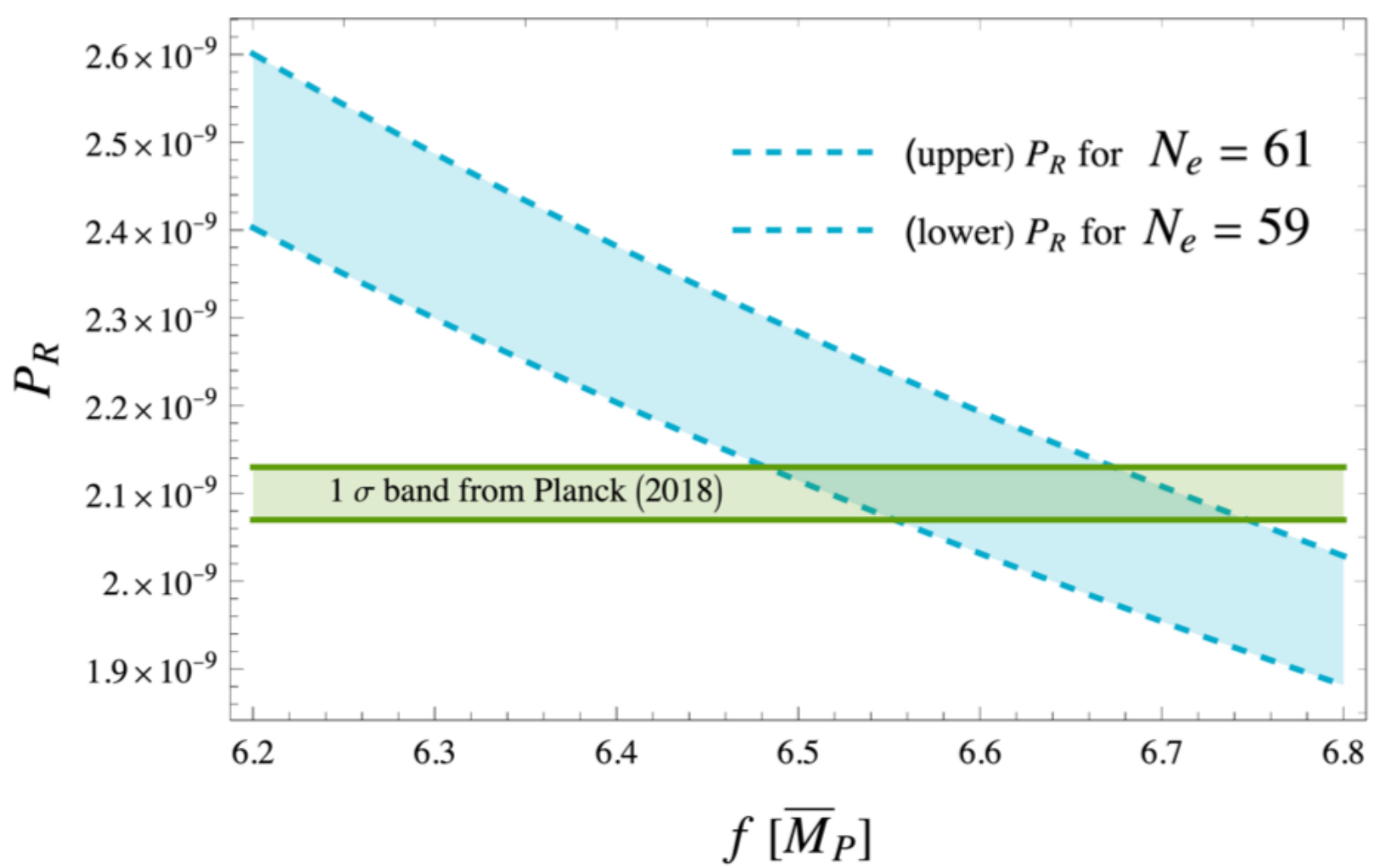} 
 \end{center} 
   \caption{\em The slow-roll parameters, $\epsilon$ and $\eta$, the spectral index $n_s$, the tensor-to-scalar ratio $r$ and the curvature power spectrum $P_R$ in natural inflation as a function of $f$. We have set $\Lambda  \approx 6 \times 10^{-3} \bp$ to fit the observed value of the curvature power spectrum $P_R$ and chosen $N=1$ and $f_2=10^{-8}$ (the value of $f_2$ influences the plot of $r$ only). The bounds from the latest Planck analysis from 2018~\cite{Akrami:2018odb} are also shown.
   }
\label{inflation-nat}
\end{figure}

  Finally the curvature power spectrum (at horizon exit) is
\be P_{R}= \frac{V_N/ \epsilon}{24\pi^2 \bp^4} \ee
and its expression for natural inflation is
\be P_{R}=\frac{f^2 \Lambda ^4 \left(1+\cos \left(\frac{N \phi}{f}\right)\right) \cot ^2\left(\frac{N \phi}{2 f}\right)}{12 \pi ^2N^2 \bp^6}. \ee

We see that the main inflationary predictions of the model can be determined analytically, like for the fundamental conformally coupled inflatons discussed before.

In Fig.~\ref{inflation-nat} we show the predictions of the model for the observables $n_s$, $r$ and $P_R$ together with the behavior of $\epsilon$ and $\eta$ to show the validity of the slow-roll approximation. In the bottom plot on the left one can clearly see that the Weyl-squared term allows perfect compatibility between natural inflation and the latest CMB data, including\footnote{Another way of achieving a viable value of $r$ in natural inflation is by adding a sizable $R^2$ term in the Palatini formalism~\cite{Antoniadis:2018yfq,Antoniadis:2018ywb}. However, as pointed out in the introduction, such term largely breaks Weyl symmetry.} the value of $r$. We also see that for $N=1$ the scale $f$ should be around $f\approx 6.6 \bp$  while $\Lambda  \approx 6 \times 10^{-3} \bp$. Since $r_E< 0.1$ the isocurvature power spectrum parameterized by $r'$ (see Eq.~(\ref{rp})) also satisfies the most recent bounds (see Fig.~7 in~\cite{Salvio:2019ewf}).

Like for the hilltop case of Sec.~\ref{model}, we have also explicitly checked for the natural-inflation implementation that the bounds in~(\ref{E2}) and~(\ref{Em}) are satisfied during the whole cosmology and, therefore, the possible runaways due to the Ostrogradsky theorem are avoided. 
 
Note that the non-minimal coupling between the Ricci scalar and $\phi$ might be generated by quantum corrections. However, it should be small, of order of the scale $N\Lambda^2/f$ at which the shift symmetry of the $\phi$ field is broken divided by the Planck mass: one has to divide by $\bp$ because the non-minimal coupling is dimensionless and because the non-minimal coupling should disappears when gravity is decoupled. This gives a negligibly small\footnote{See, however, Ref.~\cite{Ferreira:2018nav} for a discussion of non-minimal couplings in natural inflation.
} non-minimal coupling of the inflaton, that is $\xi_\phi\sim \Lambda^2/(f\bp) \sim10^{-5}$ for an order one $N$, $\Lambda  \approx 6 \times 10^{-3} \bp$ and $f\approx 6.6 \bp$.

\subsection{Asymptotically free UV-completion}\label{Asymptotically free UV-completion}

An asymptotically free UV-completion of this scenario can be obtained as follows. Take a version of QCD with a confinement scale around the Planck scale. Introduce  three flavors of tilde-quarks: $\tilde q=\{\tilde u, \tilde d, \tilde s\}$, where the tilde distinguishes from the analogous quantities in QCD (for example, $\tilde u$ is analogous to the up quark). By using the analogy with QCD, we identify $f$ with the decay constant of $\tilde\pi^0$. 
We take $f\approx 6.6 \bp$, as suggested by the results in Fig.~\ref{inflation-nat}. Like in QCD the strong dynamics forms condensates with a typical scale $\kappa \sim \langle \bar{\tilde q}' \tilde q' \rangle$ (we denote with $\tilde q'$ the Goldstone-free quark fields~\cite{Weinberg2}).   We will denote the gauge group with SU(3)$_I$ to emphasize that it is the group related to the inflaton sector. From the knowledge of QCD we obtain
\be \kappa \approx 3 f^3 \label{kappaf}. \ee
 
 Note that the tilde-condensation scale is much higher than the analogous quantity in the observed strong interactions.  This is possible because it is defined as the scale below which the asymptotically free gauge coupling  enters the non-perturbative behavior. Technically, this can be viewed as an initial condition in the renormalization group equations: the asymptotically free gauge coupling is of order $4\pi$ at the condensation scale. In the case of the observed strong interactions the experiments tell us that this scale is some hundreds of MeV, but for the new QCD-like condensation scale we do not yet have any observation fixing this value, which is, therefore, a free parameter here.

We add now, just like in QCD, quark masses such that the axial part of the global ${\rm SU(3)}_{\rm f}$  flavor group (which rotates $\{\tilde u, \tilde d, \tilde s\}$) is explicitly broken. Then we have the following spectrum of pseudo-Goldstone bosons
\bea  m^2_{\tilde K^0} &=& \frac{4\kappa}{f^2}(m_{\tilde d} + m_{\tilde s}), \label{K0} \\ 
m^2_{\tilde K^+} &=& \frac{4\kappa}{f^2}(m_{\tilde u} + m_{\tilde s}), \label{Kp} \\ 
m^2_{\tilde \pi^0} &=& m^2_{\tilde \pi^+}  = \frac{4\kappa}{f^2}(m_{\tilde u} + m_{\tilde d}),  \label{pi0}\\
m^2_{\tilde \eta^0} &=& \frac{4\kappa}{f^2}\left(\frac{m_{\tilde u}+m_{\tilde d} + 4 m_{\tilde s}}{3}\right), \label{eta0}
\eea
which can be found with EFT methods~\cite{Weinberg2}.
 Now, by choosing an inverted hierarchy $m_{\tilde  u} \gg m_{\tilde d}, m_{\tilde  s}$ we obtain that the lightest pseudo-Goldstone boson is $\tilde K^0$; this gives  just one inflaton field, $\tilde K^0$, and avoids the isocurvature bounds of Planck. Note that from~(\ref{VInf}) the mass of the inflaton is  $m_\phi =N\Lambda^2/f$ and, using the numerical values in Fig.~\ref{inflation-nat} ($\Lambda  \approx 6 \times 10^{-3} \bp$ and $N=1$ ) as well as $f\approx 6.6 \bp$, one obtains the typical value $m_\phi \approx 5.4\times 10^{-6} \bp$. 
The hierarchy $\Lambda\ll f$ corresponds to choosing the Lagrangian masses of the lightest tilde-quarks\footnote{Recall that in our setup the lightest tilde-quarks are $\tilde d$ and $\tilde s$, while the mass of $\tilde u$ is taken at a much larger scale.} $m_{\tilde q}$ to be really tiny in units of $\bp$: indeed, from $m_{\phi}^2 \sim 4  m_{\tilde q} \kappa/f^2$ (see~(\ref{K0})-(\ref{eta0})), $f\approx 6.6 \bp$, $m_\phi \approx 5.4\times 10^{-6} \bp$ and~(\ref{kappaf}) one finds
 $m_{\tilde q}  \sim m_\phi (m_{\phi}/12f) \sim 10^{-13}\bp$.

\subsection{Higgs naturalness}\label{NatNat}

The naturalness of the Higgs mass is elegantly achieved in these realizations. One can consider an asymptotically free theory with a gauge group
\be G = G_h\times SU(3)_I, \ee 
where $G_h$ is the group factor under which the Higgs is charged.

 For example, in~\cite{Pelaggi:2015kna} it was shown that the SM can be embedded in an asymptotically free theory with the trinification gauge group $G_{333}\equiv$ SU(3)$_L \times $SU(3)$_C \times $SU(3)$_R$. In this case the total gauge group will be
\be G = SU(3)_L\times SU(3)_C\times SU(3)_I\times SU(3)_R, \qquad (\mbox{trinification example}). \ee 
Other examples can be found by looking at the other asymptotically free matter sectors found in~\cite{Giudice:2014tma,Holdom:2014hla}. Even more examples can be generated by considering identical copies of these asymptotically free matter contents.

As long as the couplings of these visible sectors (charged under $G_h$) with the inflaton sector are small the Higgs mass is protected from large radiative corrections by an approximate shift symmetry acting on the Higgs field. As discussed in Sec.~\ref{The quasi-conformal scenario}, this approximate symmetry is respected by gravitational interactions if~(\ref{smallf2}) is enforced.

\subsection{Reheating}\label{Reheating}

Given that there is an asymptotically free embedding of natural inflation we study further the early-universe implications of this scenario. Here we describe how reheating can occur. 

The Lagrangian in the Einstein frame can be computed through the quantity $\Omega^2$ defined in~(\ref{footnote2b}). 
Note that the fundamental scalars $\phi_a$  undergo quantum fluctuations\cite{Enqvist:2014tta} of order $H/(2\pi)\ll \bp$ during inflation,
 therefore, assuming all of them of this order of magnitude one finds  
\be \Omega^2\sim 1-\frac{N_s}{6} \left(\frac{H}{2\pi \bp}\right)^2. \ee 
Taking now 
\be \frac{H}{2\pi}\approx 10^{13}~{\rm GeV} \label{Hinf} \ee
(having used the setup of Fig.~\ref{inflation-nat}) gives $\Omega^2 \approx 1-3\times 10^{-12} N_s$. So $\Omega^2$ is very close to one whenever $N_s\ll 10^{12}$. We assume this is satisfied. Then, as far as reheating is concerned, the Einstein-frame Lagrangian is basically equal to the Jordan frame one, but with negligibly small non-minimal couplings (see Eqs.~(\ref{footnote2}),~(\ref{footnote2b}) and~(\ref{footnote3}) and, for the matter fields with non-zero spin,  see the Einstein-frame Lagrangian in Ref.~\cite{Salvio:2018crh}). 

During inflation the inflaton energy density $\rho_I$  is very high; in particular it is much higher than the energy density of the fundamental scalars:
\be  \frac{\lambda_{abcd}}{4!} \phi_a\phi_b\phi_c\phi_d \ll 3\bp^2 H^2. \ee
However, $\rho_I$ decreases with time. When $m_\phi \gg \dot a(t)/a(t)$ the inflaton oscillates.
 Taking $m_\phi \approx 5.4\times 10^{-6} \bp$ as in the example of Fig.~\ref{inflation-nat}, the inflaton oscillations occur when the Hubble rate is say a couple of orders of magnitude below, $H_{\rm osc} \sim 5\times 10^{-8} \bp$, and the corresponding inflaton energy density is 
 \be \rho_I^{\rm osc} \approx 3 \bp^2 H_{\rm osc}^2 \sim 75 \times 10^{-16} \bp^4.\label{rhoosc} \ee To reheat the visible universe we need  eventually
\be  \frac{\lambda_{abcd}}{4!} \phi_a\phi_b\phi_c\phi_d \gg \rho_I^{\rm osc}. \label{EnDom}\ee
During inflation, as mentioned above, $\phi_a$  undergo quantum fluctuations of order $H/(2\pi)$. If $\lambda_{abcd}\ll 1$, which we assume and is typically true in asymptotically free theories, this value is kept for a long time, while $\epsilon =1$ at the end of inflation, so the Hubble rate decreases more rapidly and $\rho_I$ quickly reaches $\rho_I^{\rm osc}$.
Assuming, for example, the $\lambda_{abcd}$ to be all of the same order, $\bar \lambda$, (\ref{EnDom}) requires
\be \frac{\bar \lambda}{4!}N_s^4 \left(\frac{H}{2\pi}\right)^4 \gg \rho_I^{\rm osc}, \label{EnDom2}\ee
which can be viewed as a bound on the number of scalars: using~(\ref{Hinf}) and~(\ref{rhoosc})
\be N_s \gg \frac{10^2}{\bar\lambda^{1/4}}. \label{BoundNs}\ee 
For example, for $\bar\lambda \sim 10^{-2}$ this means\footnote{Such number of scalars can be obtained, for example, by considering several copies of the trinification model of~\cite{Pelaggi:2015kna}, which features 54 real scalars.} $N_s \gg 10^2$. One should keep in mind, however, that some of the $\lambda_{abcd}$ can be much smaller or vanishing.

When~(\ref{EnDom2}) holds, assuming the instantaneous reheating (equating the scalar to the thermal energy density), one obtains the reheating temperature 
\be T_{\rm RH} \approx \left(\frac{30\bar\lambda}{4!\pi^2 g_*}\right)^{1/4} \frac{N_s H}{2\pi}, \ee
where the number of relativistic degrees of freedom $g_*$  has to be large because of~(\ref{BoundNs}).
For the numerical values used in Fig.~\ref{inflation-nat} one has (see~(\ref{BoundNs}) and (\ref{Hinf}))
\be T_{\rm RH} \gg \left(\frac{30}{4!\pi^2 g_*}\right)^{1/4} 10^{15}~{\rm GeV}\ee 
which is typically  large  enough and in particular much larger than the electroweak scale.

Note that  what allows us to heat the universe is the presence of many weakly coupled scalars, which have sizable couplings to the observed particles and undergo large quantum fluctuations.

Finally, we observe that a coupling between the inflaton sector and the SM is always generated in this class of theories, at least through gravitational dynamics. For example, below the condensation scale gravity generates a portal coupling in the Lagrangian between the composite scalar $\phi$ representing the inflaton and the Higgs $h$ of order  $f_2^4 h^2\phi^2$.  We find that the small value of $f_2$ required by Higgs naturalness in~(\ref{smallf2}) is compatible with the lower bound on the reheating temperature derived above.

\subsection{Potential (meta)stability}

A related issue in high-energy extensions of the SM is to understand whether the electroweak vacuum is stable, unstable or metastable\footnote{We recall that the vacuum is stable if it does not decay, is metastable if its lifetime is finite but larger than the age of the universe and is unstable otherwise.}. We address here this issue in a model independent way. 

  At high enough energies the potential can be approximated by its dimension-4 part, which, for the generic class of theories considered here, reads
  \be V_4 = \frac{\lambda_{abcd}}{4!} \phi_a\phi_b\phi_c\phi_d. \ee
  In the high-energy approximation the condition of stability is $V_4> 0$  for any $\phi_a\neq 0$, such that the electroweak vacuum, $\phi_a\approx 0$, is the absolute minimum. It can be shown~\cite{Qi} that this condition is realized if and only if the $N_s$ equations
  \be \lambda_{abcd} \phi_b\phi_c\phi_d = \lambda \phi_a^3 \label{stabCond}\ee
are satisfied for non-negative $\lambda$. Eqs.~(\ref{stabCond}) can be viewed as generalized eigenvalue equations: the difference with respect to the ordinary eigenvalue equations is that here we have a fourth-order tensor $\lambda_{abcd}$ instead of a second-order tensor (that is a matrix).

The stability condition stated above gives a constraint on the values of $\lambda_{abcd}$ and applies to  {\it any} theory of the type considered here (defined at the beginning of Sec.~\ref{The quasi-conformal scenario}). However, it is useful to give a simple (but yet non-trivial) example to clarify how this constraint looks like in practise. Consider for instance an extension of the SM featuring only two scalars: so just one real scalar $\phi$ besides the Higgs $h$. The high-energy potential in this case would be
\be V_4 = \frac{\lambda_h}{4} h^4 +\frac{\lambda_\phi}4 \phi^4+\frac{\lambda_{h\phi}}4 h^2\phi^2. \ee 
In this case Condition~(\ref{stabCond}) with non-negative $\lambda$ reduces to 
\be \lambda_h >0, \qquad \lambda_\phi >0 \quad  \mbox{and} \quad 4\lambda_h \lambda_\phi -\lambda_{h\phi}^2 >0\ee
Examples of fully realistic extensions of the SM that satisfy these inequalities were found in~\cite{Salvio:2015cja}.

We also observe that total asymptotic freedom (theories where all couplings flow to zero at large energies) favours metastability: the smallness of all couplings at  high energy  suppresses the value of the potential and, therefore, raises the lifetime of the electroweak vacuum.
  
 \section{Conclusions and outlook}\label{conclusions}
 
 In this paper we explored the observable predictions of the quasi-conformal scenario, in which the dimensionless couplings in the action respect (at least approximately) Weyl symmetry. This behavior is suggested by the UV complete QG, which can hold up to infinite energy if all couplings reach a conformal fixed point at infinite energies~\cite{Salvio:2017qkx}. QG can also render the Higgs mass (technically) natural by allowing a good enough shift symmetry of the Higgs field even in the presence of gravity.
 
 We have included in the discussion the Weyl-squared term and reviewed and extended the arguments in favour of its viability in Sec.~\ref{The quasi-conformal scenario}. Basically the discussion can be addressed at two levels. At the classical level, it was shown~\cite{Salvio:2019ewf} that the possible Ostrogradsky runaways are avoided if the typical energies are below a threshold (given in~(\ref{E2}) and~(\ref{Em})), which is high enough to describe the whole cosmology. At quantum level unitarity is restored by computing probabilities by an appropriate positive norm, which is singled out 	by an experimental approach to the definition of probabilities. Moreover, the possible acausal effects are extremely diluted by the expansion of the universe.
 
 The predictions of the Weyl-squared term and the quasi-conformal non-minimal coupling of fundamental scalars have been worked out and compared with the available data of the early universe. We considered two implementations of the quasi-conformal option. 
 
 In the first one, discussed in Sec.~\ref{model}, the inflaton is identified with a fundamental scalar with a moderate ($<\sqrt{6}\bp$) field excursion, such that the effective Planck mass never becomes imaginary. An example is hilltop inflation, which agrees well with the most recent data provided by the Planck collaboration. We pointed out, however, the complexity required to UV complete models of this sort.
 
This brought us to the second implementation, which we investigated in Sec.~\ref{Natural inflation}: a pseudo-Goldstone boson as the inflaton (natural inflation), which provides a very good rationale for the flatness of the potential. In this case the non-minimal coupling can be nearly zero, rather than conformal, because a pseudo-Goldstone boson has very small non-derivative interactions. In Einstein-gravity implementations of natural inflation, the prediction of $r$ is rather large and on the verge of being excluded. On the other hand, in the presence of the Weyl-squared term, $r$ is drastically reduced for a natural Higgs mass. This can revive natural inflation. Another advantage of this realization is the possibility to be elegantly UV-completed. We have shown that an asymptotically free QCD-like sector can provide an inflaton of this type: it can be identified with the composite scalar analogous to the $K^0$ meson in ordinary strong interactions. Furthermore, we have  also shown that a satisfactory reheating can occur in this case.

We have checked, both in the hilltop and in the natural-inflation case, that the bounds~(\ref{E2}) and~(\ref{Em}) on the energies to avoid the Ostrogradsky instabilities are satisfied.

Furthermore, in both cases the main inflationary formul\ae\,\,can be analytically computed, potentially providing  simple benchmark models of inflation. 
 
 Reheating can occur in both cases too. We focused in this paper on the reheating in the UV-complete natural inflation (see Sec.~\ref{Reheating}) but it can also be realised in hilltop inflation. Indeed, one can also consider the tree-level portal coupling $\lambda_{h\phi}$ between the Higgs $h$ and the inflaton $\phi$ and the small value required by naturalness in~(\ref{NatInf}) is compatible with a high enough reheating temperature. 
 
 We take advantage of the concluding section to point out that this scenario, appropriately implemented to include a UV-complete sector containing  all SM particles, as discussed in Sec.~\ref{NatNat}, may solve all phenomenological problems of the SM. For example, dark matter, baryogenesis and neutrino masses might be due to right-handed neutrinos~\cite{Canetti:2012kh,Eijima:2018qke} (which in some models with UV fixed points are necessary~\cite{Pelaggi:2015kna}) and the electroweak vacuum stability may be achieved thanks to the extra scalars generically present in the theory (see Refs.~\cite{EliasMiro:2012ay,Salvio:2015cja,Salvio:2015jgu,Salvio:2018rv} for the stabilization through extra scalars). Therefore, we hope that the results of this paper will lead to scientific activities along these lines.

 \vspace{0.3cm}
 
\subsubsection*{Acknowledgments}

I thank G. Ballesteros and A. Strumia for useful discussions and CERN for the hospitality. 

 \vspace{1cm}
 
 \footnotesize
\begin{multicols}{2}

\end{multicols}

\end{document}